\documentclass[
 reprint,
 amsmath,amssymb,
 aps, prl,
]{revtex4-2}

\usepackage{graphicx}
\usepackage{dcolumn}
\usepackage{bm}
\usepackage{xcolor}
\usepackage{hyperref}

\begin{document}

\title{Upsurge of spontaneous knotting in polar diblock active polymers}

\author{Marin Vatin}
\affiliation{Department of Physics and Astronomy, University of Padova, Via Marzolo 8, I-35131 Padova, Italy}
\affiliation{INFN, Sezione di Padova, Via Marzolo 8, I-35131 Padova, Italy}
\author{Enzo Orlandini}
\affiliation{Department of Physics and Astronomy, University of Padova, Via Marzolo 8, I-35131 Padova, Italy}
\affiliation{INFN, Sezione di Padova, Via Marzolo 8, I-35131 Padova, Italy}
\author{Emanuele Locatelli}
\affiliation{Department of Physics and Astronomy, University of Padova, Via Marzolo 8, I-35131 Padova, Italy}
\affiliation{INFN, Sezione di Padova, Via Marzolo 8, I-35131 Padova, Italy}

\date{\today}

\begin{abstract}

Spontaneous formation of knots in long polymers at equilibrium is inevitable but becomes rare in sufficiently short chains. Here, we show that knotting and knot complexity increase by orders of magnitude in diblock polymers with a fraction $p$ of self-propelled monomers. Remarkably, this enhancement is not monotonic in $p$ with an optimal value independent of the monomer's activity. By monitoring the knot's size and position we elucidate the mechanisms of its formation, diffusion, and untying and ascribe the non-monotonic behaviour to the competition between the rate of knot formation and the knot's lifetime. These findings suggest a non-equilibrium mechanism to generate entangled filaments at the nano-scale.

\end{abstract}

\maketitle

The emergence of topological signatures is ubiquitous in soft matter physics, ranging from conventional polymers, bio-polymers, defect loops, and vortices~\cite{tubiana2024topology}. Knots are the best-known example of a topological state due to their practical relevance in everyday life. Still, they are also observed at the micro and nano-scales, where their presence can have a significant impact on the physical properties of the hosting system~\cite{sumners1988knots,saitta1999influence,arai1999tying,katritch1996geometry,orlandini2008slow,crisona1999topological,orlandini2007statistical,marenduzzo2010biopolymer,kanaeda2009universality,rosa2011structure}. 
In biology, knots in DNA can affect the regulation of gene expression~\cite{portugal1996t7,deibler2007hin,liu2009and}, the process of DNA replication/recombination~\cite{wasserman1986biochemical} and the spatial organisation and ejection dynamics of viral DNAs~\cite{marenduzzo2009dna,matthews2009knot,marenduzzo2013topological}. Knotted motifs have been observed in proteins and they are believed to play a crucial role in the folding and mechanical stability of the polypetides~\cite{taylor2000deeply,hsu2023folding,virnau2006intricate,sulkowska2012conservation,soler2013effects,jackson2017fold}.
{Going beyond thermodynamic equilibrium, topological signatures}
have been recently explored in active systems such as the cytoskeleton~\cite{fletcher2010cell}, actomyosin networks~\cite{koenderink2009active}, gliding assays~\cite{sciortino2023polarity}, chromatin~\cite{weber2012nonthermal,zidovska2013micron} worms assemblies~\cite{patil2023ultrafast, deblais2023worm} and active nematics~\cite{doostmohammadi2018active}. 
Natural playgrounds to explore {the physics of these systems} are the so-called active polymers, that lately have attracted interest because of their unprecedented statistical properties and wide range of applications~\cite{winkler2020physics, zhu2024non}. 
The relevance of {the interplay between the} activity 
{and the topology} of fluctuating filaments has emerged 
in systems of active linear chains under confinement\cite{manna2019emergent, das2019dynamics} in melts~\cite{tejedor2019reptation, tejedor2023molecular, breoni2023giant, ubertini2024universal} as well as in diluted~\cite{locatelli2021activity, theeyancheri2022migration, lamura2024excluded} and concentrated~\cite{Active_topoglass_NatComm20, miranda2023self,micheletti2024topology} solutions of active rings.\\ 
{These works focus on unknotted rings or mutual entanglement in linear chains; self-entanglements have been scarcely investigated in this context. In Ref.~\cite{foglino2019non}, the authors observed fewer knots than in equilibrium in a coarse-grained active polymer model with explicit motors; recently, it was shown that a grafted polar active polymer spontaneously formed knots~\cite{li2024activity}.} 
{Indeed, in these out-of-equilibrium systems, one may argue that the local stresses due to activity may alter the mechanism of entanglement formation and, consequently, the frequency of spontaneous knot formation. As such, they represent an intriguing venue for producing knotted filaments at the nano-scales.} \\
In this work, we {study the formation of self-entanglements in} an active/passive diblock copolymer where only a fraction of the monomers, $p$, are active.  Focusing on the knotting probability and the knot complexity in steady state we show that, for relatively short chains, the likelihood of observing a knot increases by orders of magnitude if compared with the equilibrium case. Strikingly, the knotting probability is a non-monotonic function of $p$ with an optimal value that seems independent of the monomer's activity.  
By exploring the knotting/unknotting events and the knot motion along the system we ascribe the non-monotonic behaviour to the sum of two competing effects: the rate of knot formation, always occurring at the active extremity, and the knots' lifetime, dominated by its residence time in the passive region.
The system can therefore be steered toward forming more knots by tuning the fraction $p$.\\ 
\begin{figure*}[t]
	\centering
	\includegraphics[width=0.9\textwidth]{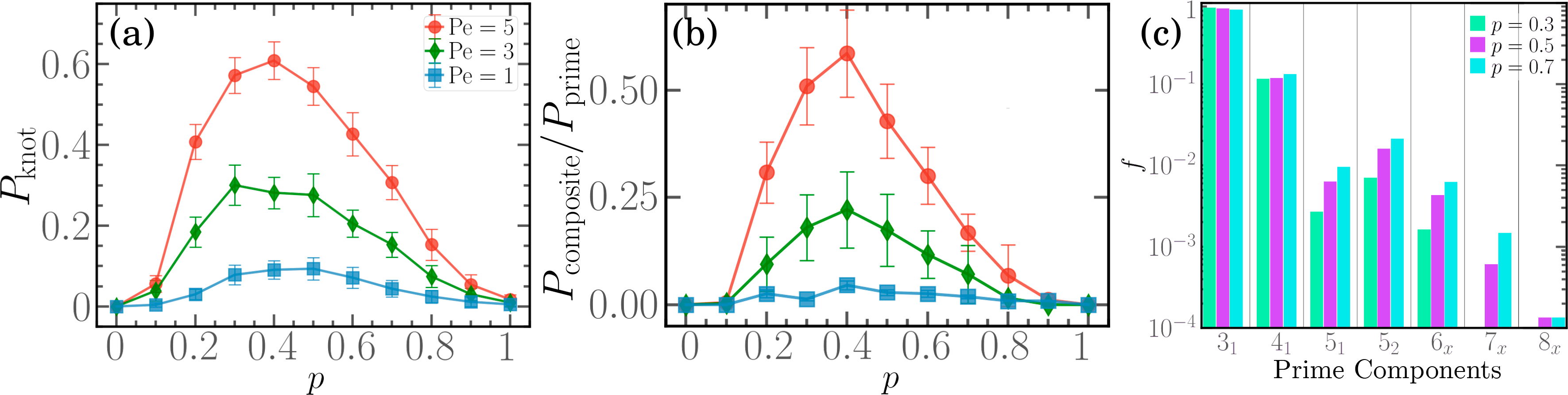}
	\caption{(a) Knotting probability, (b) ratio of occurrences of composite knots over prime knots, as a function of $p$ for $N = 300$, $\mathrm{Pe} = 1$ (blue squares), $\mathrm{Pe} = 3$ (green diamonds), $\mathrm{Pe} = 5$ (red circles). Error bars refer to the standard error of the mean.
    (c) Fraction of the different prime components for $N = 300$, $\mathrm{Pe} = 5$, $p =$ 0.3 (green), 0.5 (purple), 0.7 (cyan).
	}
	\label{fig:Fig_1}
\end{figure*}
We consider a model of a flexible linear polymer chain of length $L=N\sigma$, $\sigma=1$ being the bead's diameter; we focus on $N=300$, a length that in passive polymers is too short to observe knot formation events~\cite{tubiana2013spontaneous}. Following Ref.~\cite{vatin2024conformation}, we design the active/passive heterogeneity as a diblock copolymer, namely a chain made by two contiguous blocks, the active one of length $p\sigma$ and the passive one of length $(N-p)\sigma$.
The active block is located at one of the two ends of the chain, which we call the ``head'' of the polymer.  Each active monomer is self-propelled by a force $\mathbf{f}_a$, with constant magnitude ${f}_a$, directed as the local tangent; this is known as tangential or \emph{polar} self-propulsion. The overall active force points in the direction of the head (also referred to as the ``leading'' end)~\cite{vatin2024conformation}. 
Following~\cite{bianco2018globulelike,vatin2024conformation}, both end beads are made passive (see a sketch of the model in Fig. S1).
We characterise the strength of the self-propulsion via the P\'eclet number, $\mathrm{Pe} = f_a \sigma / k_B T$, $k_B T$ being the unit of energy~\cite{bianco2018globulelike, vatin2024conformation}. The system is evolved via Langevin dynamics simulations 
~\cite{kremer1990dynamics}
(see Suppl. Mat. Sec.~1A,~B~\footnote{See Supplemental Materials at [URL will be inserted by publisher] for additional definitions and theoretical details, model and simulation details, and additional numerical data, including Refs.[70-72].}).
To monitor the formation, motion, and disappearance of a physical knot along the chain,  we employed Kymoknot~\cite{tubiana2018kymoknot} a software that uses a minimally interfering scheme to close an open chain (see  Suppl. Mat. Sec.~2).  This procedure allows us to assign a topological state (knot type) to an open chain and locate the knotted portion. {Employing the standard concept of crossing number $n_c$ ~\cite{tubiana2024topology},} we limit ourselves to knotted configurations with  $n_c\le 8$~\footnote{In our analysis, we found almost no knots with $n_c>$8.}. Moreover, since the knot identification is based on the Alexander polynomial $\Delta$, the composite knots are distinguished, by visual inspection, from the corresponding prime knots with $n_c=8$ having the same $\Delta$ (e.g. the composite knots  $3_1\#3_1$ and $3_1\# 4_1$)~\footnote{Knots that can be factorised into simpler knots are called \emph{composite} knots.  If a knot cannot be factorised the knot is a \emph{prime knot}}.
{To improve the statistical significance of the results, we simulate $M=$ 100 independent trajectories for each pair $(p,~\mathrm{Pe})$ considered. We assess the conformational steady state condition of the polymer by looking at the time-dependence of its gyration radius (see Suppl. Mat. Fig.~S3); at steady state knots form, diffuse, and eventually untie. We thus sample the simulated trajectories in the steady state every $\tau_s = 10 \tau$, with $\tau$ the unit of time (see Suppl. Mat. Sec. 1B). }\\
In Fig.~\ref{fig:Fig_1}A we report the probability $P_{\mathrm{knot}}$ that, at steady state, the system displays a knot as a function of $p$. It is worth noting that: (i) At fixed $p$, $P_{\mathrm{knot}}$ increases monotonically with $\mathrm{Pe}$, at least in the range of the Peclet number considered here~\footnote{For larger values of the Peclet number (e.g. $\mathrm{Pe}=$20), a considerable fraction of configurations are compact~\cite{vatin2024conformation}. Since these conformations are long-lived and highly condensed, knotting/unknotting events are rare. As such, large $\mathrm{Pe}$ values are avoided.}. Moreover, for intermediate values of $p$,  $P_{\mathrm{knot}}$ is orders of magnitudes higher than the one measured in the fully passive ($p$=0) and fully active ($p$=1) case. Indeed, from previous studies~\cite{tubiana2013spontaneous} one can extrapolate  $P_{\mathrm{knot}} \approx 4\cdot 10^{-4}$ for passive polymers of contour length $L=300\sigma$, implying an enhancement of the order of $~10^2-10^3$. (ii) Notably, for all activities investigated $P_{\mathrm{knot}}$ is \emph{non-monotonic} in $p$ reaching a maximum $p^*\approx 0.4$ that seems to be insensitive  to the activity. Finally, at $p=1$ (fully active polymers), $P_{\mathrm{knot}}$ is negligible, in qualitative agreement with Ref.~\cite{foglino2019non}.\\
Next, we investigate the topological complexity of the observed knots by partitioning the knot population according to their knot type (knot spectrum). Specifically, we look at the ratio between the probability of detecting either composite or prime knots as a function of $p$ (Fig.~\ref{fig:Fig_1}b).  
Interestingly, {the ratio is still non-monotonic in $p$; further,} while composite knots become more abundant upon increasing $\mathrm{Pe}$, most knots are prime knots. Their dominance is {a feature of this system, that sets it further apart from the passive case where, for entropic reasons, the knot spectrum of very long polymers is} dominated by the connected sum of several prime knots~\cite{orlandini2007statistical}. 
If we restrict the analysis of the knot type to all prime knots ({including the prime components} in composite knots), we observe that (i) there is an overwhelming majority of simple prime knots ($3_1$ and $4_1$); (ii) the knot complexity mildly increases with $p$; 
(iii) the knot spectrum broadens upon increasing $\mathrm{Pe}$ (see Suppl. Mat. Fig.~S.6) and (iv) the probability of observing a $5_2$ knot (twist knot)  is always higher than the one of a $5_1$ knot (torus knot). This holds for passive polymers in the swollen phase but also in everyday life, since twist knots are easier to form accidentally~\cite{adams1994knot}.
It is important to stress that the dramatic increase in spontaneous knotting is observed for swollen, relatively short chains. This makes the active diblock copolymers 
a valid alternative 
to approaches based either on polymer compression ~\cite{amin2018nanofluidic, rothorl2022knot} or electrohydrodynamic instability\cite{tang2011compression, klotz2017dynamics}; these techniques may provide better performances in knot production but tend to yield very complex knots due to the relatively high degree of compaction reached by the polymer substrate~\cite{michieletto2020separation, rothorl2022knot, klotz2017dynamics, tubiana2024topology}.\\
\begin{figure*}[t]
	\centering
	\includegraphics[width=0.85\textwidth]{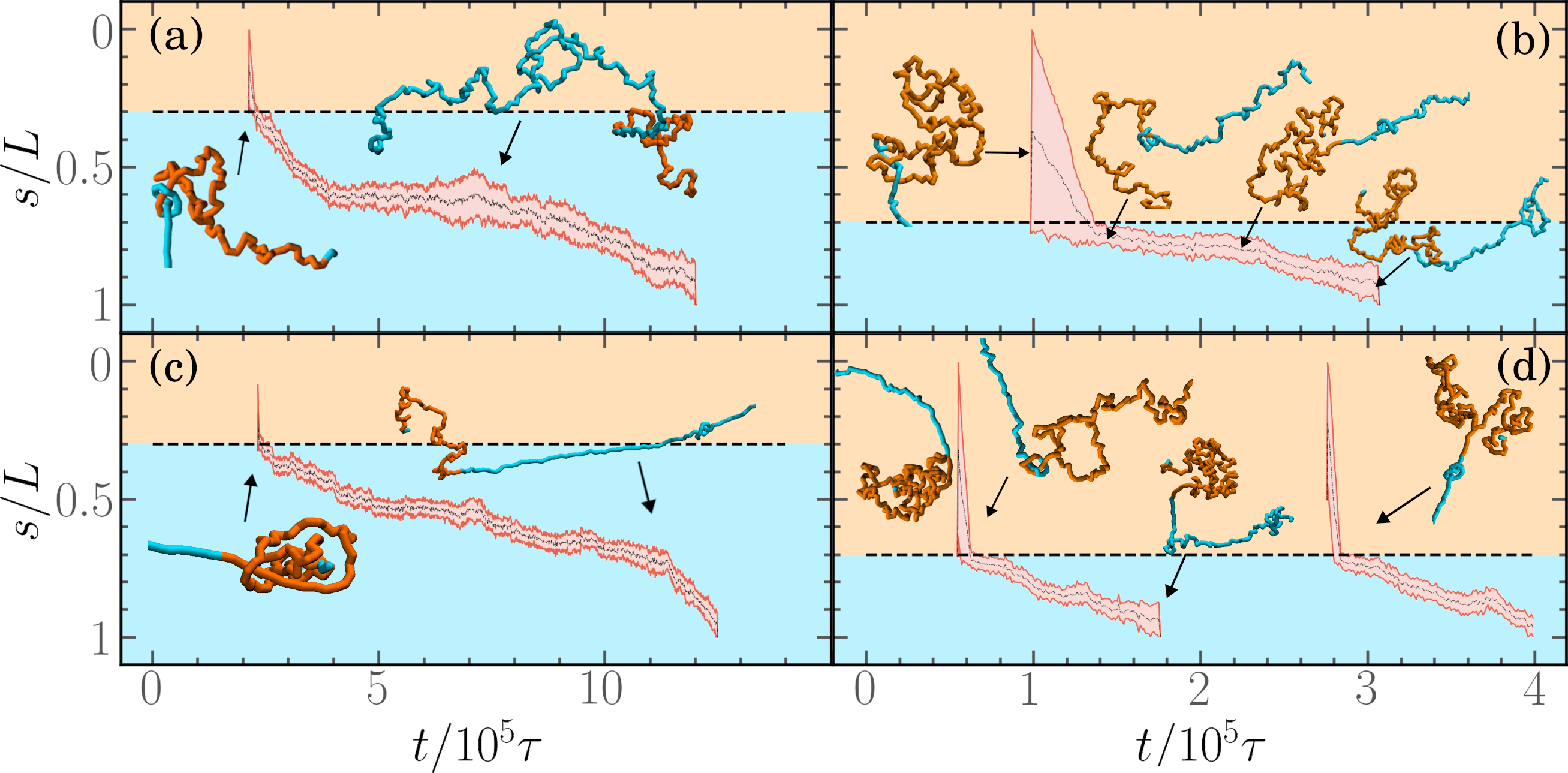}
	\caption{Kymograph of the reduced position of the ends of the knotted segment (red lines) and the center of the knot (full black line) for (a) $p = 0.3$ and $\mathrm{Pe} = 1$, (b) $p = 0.3$ and $\mathrm{Pe} = 5$, (c) $p = 0.7$ and $\mathrm{Pe} = 1$, (d) $p = 0.7$ and $\mathrm{Pe} = 5$. 
    Black dashed lines mark the boundary between the active and passive sections. 
    {Active blocks are represented in orange and passive ones in blue.}
	}
	\label{fig:histories_ends}
\end{figure*}
To understand the non-monotonic behaviour of $P_{\mathrm{knot}}$, 
we looked at the dynamics of knot formation, diffusion, and disappearance. In the kymographs of Fig.~\ref{fig:histories_ends} we show examples of the time evolution of the reduced position and size of a knot for active diblock copolymers at $\mathrm{Pe}=1$, $p=0.3,~0.5$ (panels a,~b) and $\mathrm{Pe}=5$, $p=0.3,~0.5$ (panels c,~d). 
Notice that knots always form in the active region. The reason is twofold: first, the tangential activity dominates the dynamics of the whole chain, also modifying the conformation of both blocks~\cite{vatin2024conformation}. In particular, the passive block is stretched by the active force; {as such, the back-folding of the passive free end is unfavoured, hindering knot formation. Additionally}, the active block, if sufficiently long, attains a relatively compact, globule-like conformation with an activity-induced effective rigidity at short scales~\cite{bianco2018globulelike}, further characterized by the formation of ``loop''-like structures~\cite{foglino2019non, locatelli2021activity, ubertini2024universal}. {Secondly, active monomers possess an enhanced mobility, that allows the free end (the ``head'') to increase the chance of a successful back-folding event within the globule-like region.} 
{Once formed, the knot travels from the head to the tail. Indeed, in the active region, the knot undergoes a sort of ``railway motion'', because the conformation of a polar active linear polymer follows the trajectory set by the leading end. This motion drives the knot towards the passive block that is kept under tension by the active force.} Notably, the knotted portion, often initially spread (or de-localized) over the whole active block, localizes around the boundary upon when reaching the passive region. This is mainly due to the mobility difference between the passive and the active blocks, {as one of the two edges of the knotted section reaches the boundary first and becomes much less mobile than the other. After crossing said boundary, the knot remains quite tight, again because of the activity-induced tension, and keeps cruising down along the contour.}\\Notice that the migration speed in the active region {is roughly} proportional to the ``active'' velocity of the monomers $v_a=f_a/\gamma$, $\gamma$ being the friction coefficient. For example, for $p=0.7$, the time the knot spends in the active section goes from $\approx 5\cdot 10^4 \tau$ at $\mathrm{Pe}=$1 to $\approx 10^4 \tau$ at $\mathrm{Pe}=5$ (see Fig.~\ref{fig:histories_ends}b,~d). Interestingly, the same quantity in the passive section is not proportional to the activity. Indeed, upon increasing $\mathrm{Pe}$ by a factor of five, the migration time does not decrease accordingly (see again Fig.~\ref{fig:histories_ends}).\\
\begin{figure}[t]
	\centering
	\includegraphics[width=0.8\columnwidth]{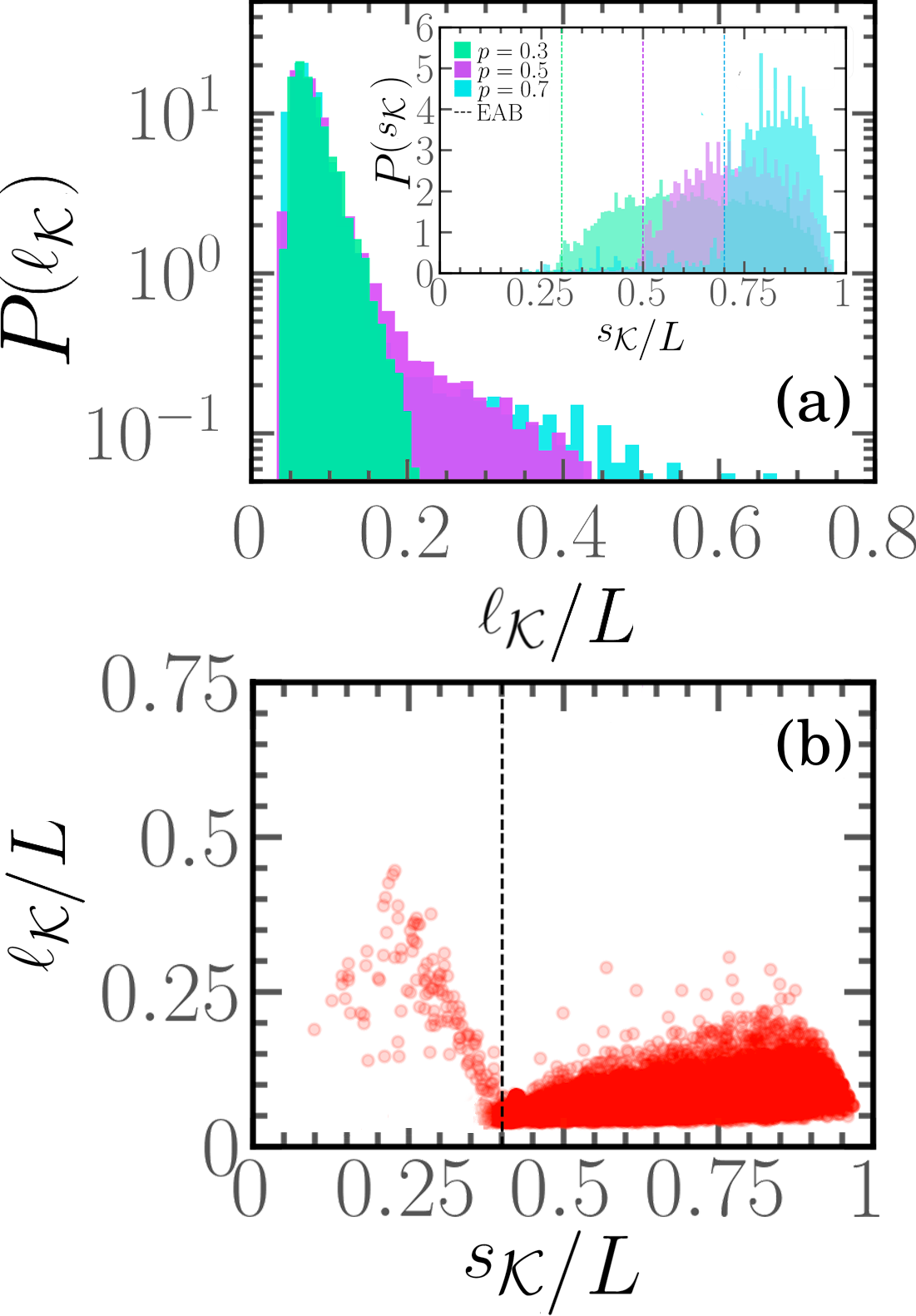}
    \caption{
	(a) Distribution of the knot length $P(\ell_{\mathcal{K}})$ and (inset) of the position of the knots' centre $P(s_{\mathcal{K}})$ for $p = $0.3 (green), 0.5 (purple), 0.7 (cyan). 
    (b) Scatter plot of $\ell_{\mathcal{K}}/L$ and  $s_{\mathcal{K}}/L$ for $p=0.4$. In both panels, $N = 300$, $\mathrm{Pe} = 5$.
	}
	\label{fig:Fig_2}
\end{figure}
Further, the relative amount of active beads $p$ also influences the length of the knotted portion $\ell_{\mathcal{K}}$ as visible in Fig.~\ref{fig:Fig_2}a, reporting the knot length distribution, $P(\ell_{\mathcal{K}})$, for different values of $p$ at fixed $\mathrm{Pe}=5$ (see also Suppl. Fig.~S7 for additional values of $p$ and $\mathrm{Pe}$). Indeed, while the most probable value of the distribution does not change with $p$, a substantial change in the distribution's shape is observed. At low values of $p$, $P(\ell_{\mathcal{K}})$ is characterised by a single exponential decay in $\ell_{\mathcal{K}}$; upon increasing $p$, the distribution at large $\ell_{\mathcal{K}}$ crosses over a second exponential behavior with a much smaller decay constant. Eventually, at $p=1$, when the chain is fully active, the knots de-localize over the whole polymer (see Suppl. Mat. Fig.~S8). 
This behaviour reflects the phenomenology described previously. Knots in the active region are de-localized; as such, the probability of observing knots with a large size increases with $p$. However, these large knots remain rare as they localize quickly. So, it is much more probable to observe a tight knot, residing in the passive region; {this can be seen in the distribution of $s_{\mathcal{K}}$, the position of the centre of mass of the knot along the contour (inset of Fig.~\ref{fig:Fig_2}a): the probability of observing the knot in the active region is small and, unless $p$ is sufficiently large, occurrences deep in the active block are rare (see also Suppl. Mat. Fig.7,~9,~10). This is confirmed in Fig.~\ref{fig:Fig_2}b, where we report the scatter plot of $\ell_{\mathcal{K}}$ as a function of $s_{\mathcal{K}}$: in the active region knots are rare and quite de-localized, while they are abundant and rather compact in the passive block.} {These features seem to be tied to the diblock copolymer design: a similar phenomenology} can be observed qualitatively for a knot in a diblock copolymer ring (see Suppl. Mat. Fig.~{S5}). 
\begin{figure}[b]
	\centering
	\includegraphics[width=0.75\columnwidth]{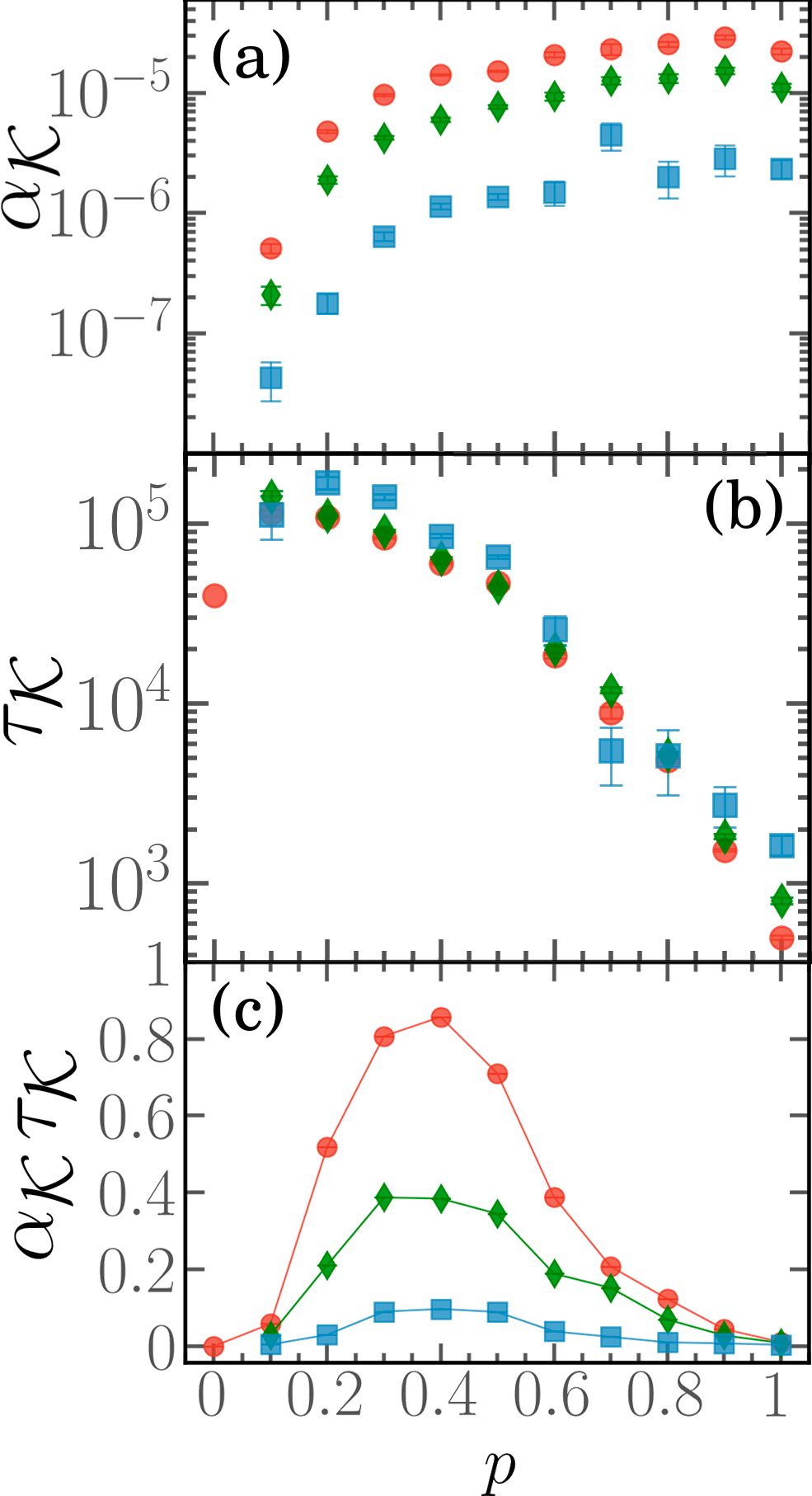}
	\caption{(a) Knot formation rate $\alpha_{\mathcal{K}}$, (b) knot lifetime $\tau_{\mathcal{K}}$ and (c) product $\alpha_{\mathcal{K}}\, \tau_{\mathcal{K}}$ as a function of $p$ for $N = 300$ and, $\mathrm{Pe} = 1$ (blue squares), $\mathrm{Pe} = 3$ (green diamonds), $\mathrm{Pe} = 5$ (red circles).
	}
	\label{fig:rateofformationlifetime}
\end{figure}
The above findings allow us to rationalize the non-monotonicity of the knotting probability (Fig.~\ref{fig:Fig_1}a) as follows: knots are created by the action of the leading end, that backfolds in the globule-like conformation of the active block. This suggests that at fixed $\mathrm{Pe}$ the rate of knot formation should increase upon increasing $p$, as the region where self-entanglement can easily form increases. This picture is confirmed in Fig.~\ref{fig:rateofformationlifetime}a, where we show that the rate of knot formation  $\alpha_{\mathcal{K}}$ (see also SM, Sec.~3) steadily increases with $p$ and $\mathrm{Pe}$.\\Once formed, the knot starts migrating from the active to the passive end, where eventually it unties. The average knot lifetime, $\tau_{\mathcal{K}}$, i.e. the period when the knot can be observed on the chain, is estimated in Fig.~\ref{fig:rateofformationlifetime}b (see Suppl. Mat. Sec.~3). Interestingly, the knot lifetime decreases as a power law with $p$ but depends weakly on $\mathrm{Pe}$, except for $p=1$. This is consistent with the observation that the knot spends most of its lifetime in the passive region, where its migration speed is roughly independent of $\mathrm{Pe}$.\\ 
The non-monotonic behavior of the product $\alpha_{\mathcal{K}}\tau_{\mathcal{K}}$, reported in Fig.~\ref{fig:rateofformationlifetime}c, confirms that the competition between the knot formation rate (in the active region) and its residence time (mostly in the passive region) rationalizes the non-monotonic behavior of $P_{\mathrm{knot}}$.\\ 
In summary, we studied the spontaneous knot formation in the steady state of polar active diblock copolymers, reporting a remarkable enhancement in the knotting probability $P_{\mathrm{knot}}$, compared to the standard passive case. This probability is non-monotonic in $p$, suggesting a specific 
value of $p$ for optimal yield. We characterized the formed knots in terms of their topological complexity and size revealing that they are mostly simple and subject to a tightening process as they migrate from the active to the passive region.
Finally, we rationalized the observed non-monotonicity as the result of the interplay of the knots' formation rate and lifetime. Interestingly, self-entanglement is, effectively, irrelevant for fully active polymers as knots are ephemeral and have no effect on their dynamics.\\
We thus show that polar active diblock copolymers can be a viable substrate for knot production with high yield. 
The polymers considered fluctuate in bulk and the knots can eventually dissolve. However, since activity can be turned on or off by external fields~\cite{chen2014janus,zhu2019propulsion, diwakar2022ac} and current state-of-the-art experimental techniques allow for precise single filament manipulation~\cite{klotz2020experimental}, we argue that knots, once created can be trapped and stored for later use. 
The proposed architecture may be useful not only to study knots but also to improve the fabrication process of microarchitected topological materials, that have been shown to possess remarkable properties, in terms of robustness and compliance~\cite{moestopo2020pushing, moestopo2023knots}.

\begin{acknowledgments}
The authors thank M. Baiesi for the helpful discussions. 
The authors would like to acknowledge the contribution of the COST Action CA17139 (\href{https://eutopia.unitn.eu/}{eutopia.unitn.eu}) funded by COST (\href{https://www.cost.eu/}{www.cost.eu}). This work has been funded by the project ``LOCA\_BIRD2222\_01'' of the University of Padova. E.L. acknowledges support from the MIUR grant Rita Levi Montalcini.E.O. acknowledges support from grant PRIN 2022R8YXMR funded by the Italian Ministry of University and Research.  The authors acknowledge the CINECA award under the ISCRA initiative, for the availability of high-performance computing resources and support. The authors acknowledge CloudVeneto for the use of computing and storage facilities.
\end{acknowledgments}


%

\newpage

\title{}
\widetext
\clearpage
\begin{center}
\textbf{\Large  \\ \vspace*{1.5mm} Upsurge of spontaneous knotting in polar diblock active polymers - Supplemental Material} \\
\vspace*{5mm}
Marin Vatin, Enzo Orlandini, Emanuele Locatelli
\vspace*{10mm}
\end{center}

\onecolumngrid

%

\setcounter{equation}{0}
\setcounter{figure}{0}
\setcounter{table}{0}
\setcounter{page}{1}
\setcounter{section}{0}
\setcounter{page}{1}
\makeatletter
\renewcommand{\theequation}{S\arabic{equation}}
\renewcommand{\thefigure}{S\arabic{figure}}
\renewcommand{\thetable}{S\arabic{table}}
\renewcommand{\thesection}{S\arabic{section}}
\renewcommand{\thepage}{S\arabic{page}}

\onecolumngrid

\section{Model and simulation details}

\subsection{Active polymer model}
\label{sec:model}

Polymer chains are modeled as fully flexible, self-avoiding bead-spring linear chains consisting of $N=300$ monomers, suspended in a 3D bulk fluid. Self-avoidance between any pair of monomers is implemented via a truncated and shifted Lennard-Jones (LJ) potential: 
\begin{equation}
V_{\text{LJ}}(r) =
\begin{cases}
  4\epsilon \left[ \left(\frac{\sigma}{r} \right)^{12}- \left(\frac{\sigma}{r}\right)^{6}+\frac{1}{4}\right] & \text{for~}  r < 2^{1/6}\sigma \\
  0 & \text{for~} r \geq 2^{1/6}\sigma
\end{cases}
\end{equation}
where $\sigma=1$ is the diameter of the monomer and is taken as the unit of length, $\epsilon=10\,k_B T$ with $k_B T$ the scale of energy and $r=|\vec{r}_i - \vec{r}_j|$ is the Euclidean distance between the monomers $i$ and $j$ positioned at $\vec{r}_i$ and $\vec{r}_j$, respectively. 
The Finitely Extensible Nonlinear Elastic (FENE) potential~\cite{kremer1990dynamics}
\begin{equation}
	V_\mathrm{FENE}(r) = -\frac{K r_0^2}{2} \ln \left[ 1 - \left( \frac{r}{r_\mathrm{0}} \right)^2 \right]
\end{equation}
acts between any pair of consecutive monomers along the polymer chain. We set $K= 30\,\epsilon/\sigma^2=300\,k_\mathrm{B}T/\sigma^2$ and $r_0=1.5\sigma$ in order to avoid strand crossings.

Activity is introduced as a tangential self-propulsion\cite{bianco2018globulelike}. An active monomer $i$ at position $\vec{r}_{i}$ is subject to an active force $\vec{f}^{\mathrm{a}}_{i} = f^{\mathrm{a}} \hat{t}_i$ where $\hat{t}_i = (\vec{r}_{i+1} - \vec{r}_{i-1})/|\vec{r}_{i+1} - \vec{r}_{i-1}|$ is the normalized tangent vector. The end monomers are always passive. The strength of the activity is controlled, by varying the dimensionless parameter $\mathrm{Pe}=|{f_\mathrm{a}}|\sigma/k_\mathrm{B}T$ called the P\'eclet number.\\
\begin{figure}[h!]
\includegraphics[width=0.7\textwidth]{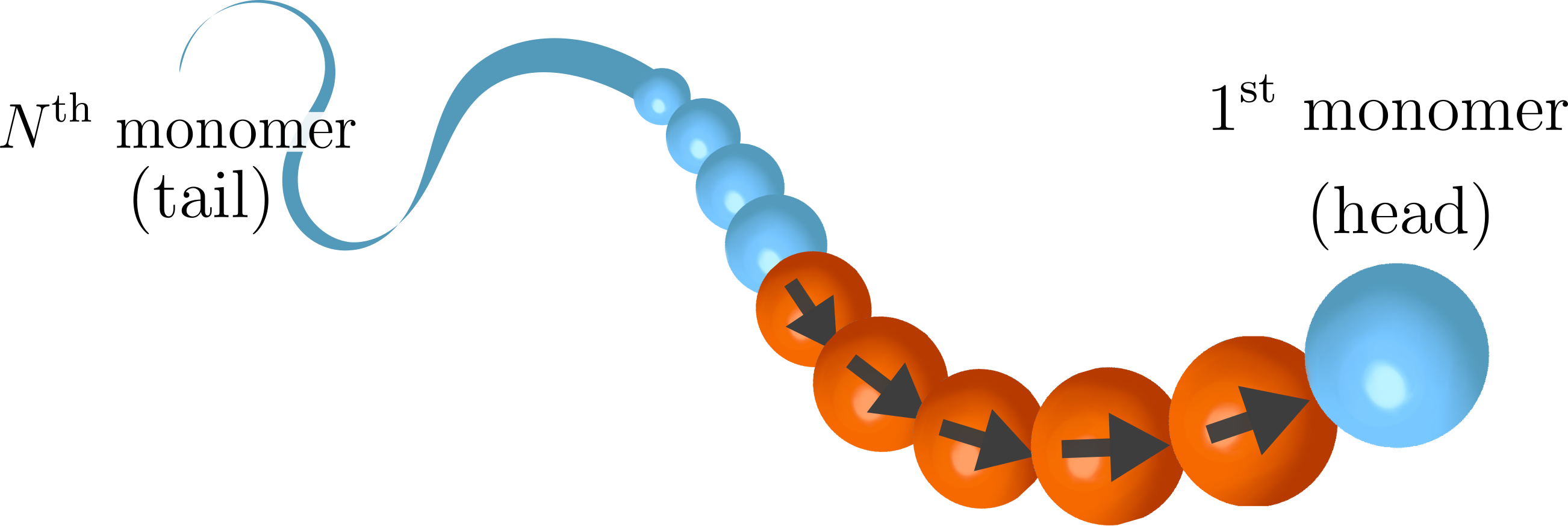}
\caption{Sketch of the diblock copolymer architecture. Active monomers are in orange, the arrows indicating the direction of the self-propulsion, passive monomers are in blue.}
\label{fig:sketch}
\end{figure}
We consider the special case in which only a fraction $p$ of the chain is composed by active monomers. The $N\cdot p$ monomers are grouped in a contiguous block, placed at one of the free end of the chain; since the last monomer is always passive, the active block starts from the nearest neighbor of the chosen end monomer. We are therefore considering di-block polymers composed of two blocks, one passive and one active; in the special cases $p=0$ and $p=1$ we obtain passive and fully active polymers, respectively. 

\subsection{Simulation details}
\label{sec:sim}	


The polymer chains, initially unknotted, are simulated in bulk conditions, by means of periodic boundary conditions, employing the open source code LAMMPS\cite{thompson2022lammps}, with in-house modifications to implement the tangential activity. Langevin Dynamics simulations are performed, disregarding hydrodynamic interactions. The equations of motion are integrated using the velocity Verlet algorithm, with an elementary time step $\Delta t = 10^{-3}$. The unit of mass, length, and energy are set to $m=1$, $\sigma=1$, and $k_B T = 1$, respectively. The unit of time is $\tau= \sqrt{m \sigma^2/ k_B T} = 1$. The overdamped regime is ensured by setting the friction coefficient $\gamma=20 \tau^{-1}$\cite{fazelzadeh2022effects}.


Polymers of length $N = 300$ are simulated for different percentages of active sites $0.0 \leq p \leq 1.0$,
with different values of the activity parameters: $\mathrm{Pe} =$ 1,3,5. $M=$100 independent trajectories are simulated for each set of parameters. After reaching a steady state, production runs are performed for 
$2 \times 10^6$ time steps and polymer conformations are sampled at a rate of $10^4$ time steps, that is larger than the decorrelation time of the end-to-end vector\cite{vatin2024conformation}. Simulations for all values of $\mathrm{Pe}$ and $p \geq 0.5$ have been extended by $14 \times 10^4$ time steps with a sampling rate of $10^2$ time steps in order to collect data regarding the short time dynamics of the knots.

\section{Knot analysis}
\subsection{Knot detection algorithm}
\label{sec:detect_algo}

Identifications of the knots are performed using the KymoKnot algorithm\cite{tubiana2018kymoknot}. This algorithm is based on the computation of standard topological invariants; for open filaments, a closing procedure of the linear backbone with auxiliary arcs is performed. More precisely, Alexander polynomial are used as topological invariants \cite{orlandini2007statistical} and the chains are closed following a "minimally interfering closure" strategy: the ends are connected directly if they are sufficiently close to each other otherwise are connected to the convex hull of the chain\cite{tubiana2011probing}.\\
%
We perform KymoKnot's analysis on each frame of the $M=$100 independent trajectories. If, considering the whole chain, a composite knot or a very complex prime knot is detected, we run Kymoknot on a portion of the chain, centered at the position of the detected knot. The size of such a portion is enlarged iteratively, until a prime knot is detected. If the starting result was a complex prime knot and such result is confirmed, we terminate the search; otherwise, the process is repeated on a different subportion until all the prime knots, hosted within the portion of the chain that contains the composite knot, are detected.
This process ensures that the algorithm is able to detect all the prime knots present on the chain. Indeed, in certain conditions, the knot lifetime is larger than the average time lag between two knotting events. In these conditions, this analysis allows us to distinguish and track knots that are effectively distinct and non-interacting and to correctly compute their lifetime.


\subsection{Steady state assessment: knotting frequency and mean gyration radius}

\begin{figure}[h!]
\includegraphics[width=0.8\textwidth]{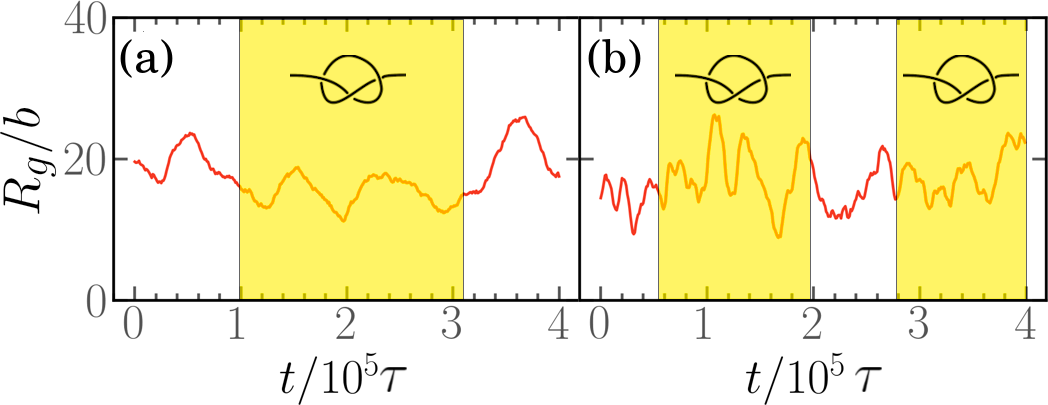}
\caption{Example of the evolution of the gyration radius $R_g$ of the chains for a) $\mathrm{Pe}=1$ and $p=0.7$ and b) $\mathrm{Pe}=5$ and $p=0.3$ presented in the main text. Yellow rectangles highlight the time intervals when the chains are knotted.}
\label{fig:rg_two}
\end{figure}
In this section, we briefly show how we checked the steady state in our simulations. We assess steady state by means of two observables: a metric one, the gyration radius and a topological one, the fraction of knotted chains. We do not rely only on the metric property because, in these systems, knots do not distinctively change the gyration radius of the chain; two examples are shown in Fig.~\ref{fig:rg_two}. As shown, there is not an evident variation in the gyration radius when knots are present. Indeed, the conformations attained by the chains are subjected to large fluctuations, especially at large values of $\mathrm{Pe}$. Further, knots are either de-localized in a region where the polymer is relatively compact or are localized; so, in both cases, they do not affect the gyration radius considerably.\\   
\begin{figure}[h!]
\includegraphics[width=0.8\textwidth]{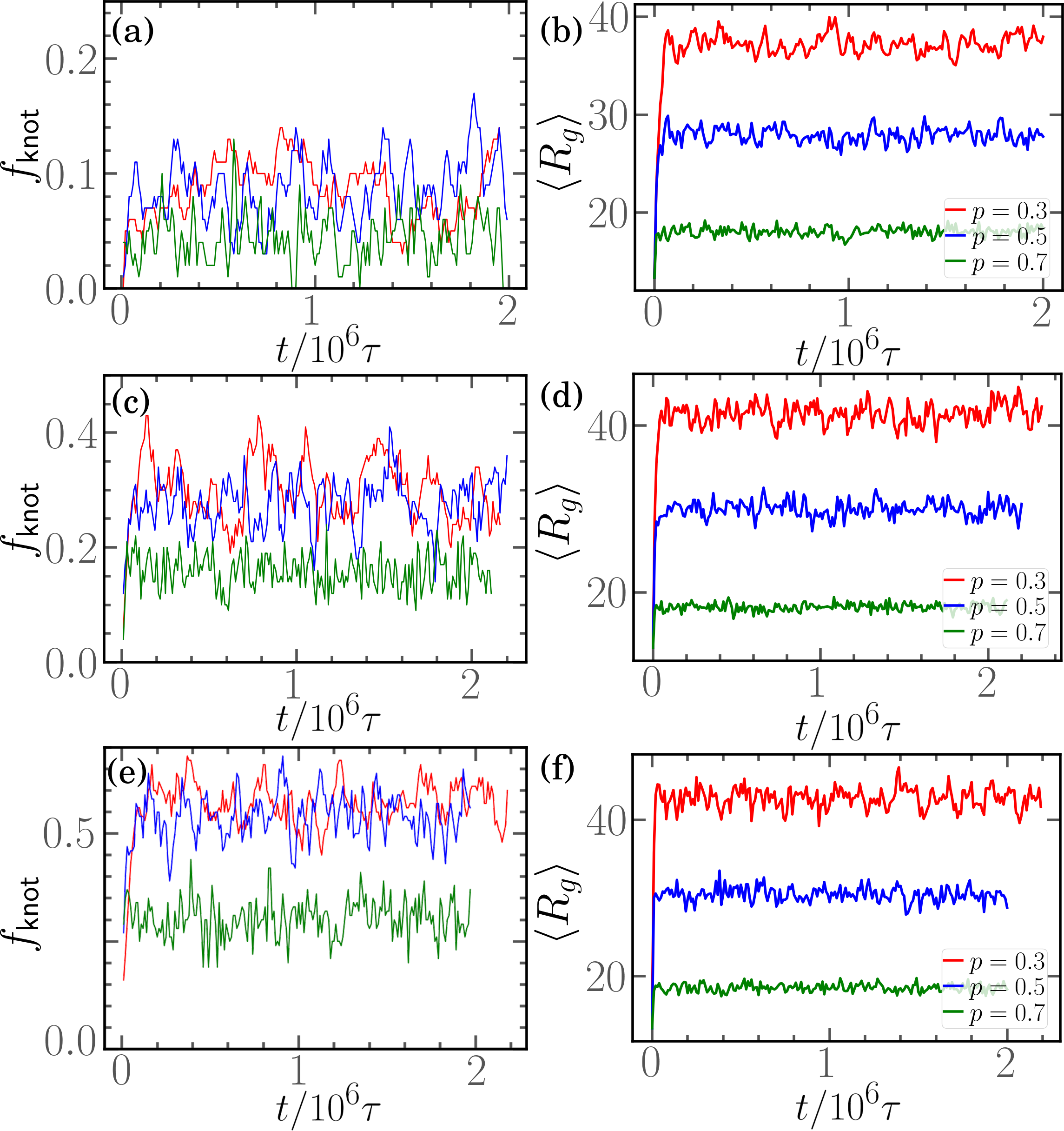}
\caption{(a,c,d) Knotting frequency (i.e. fraction of knotted polymers) and (b,d,f) mean gyration radius $\langle R_g \rangle$ as a function of time for $N=300$, $p=0.3, 0.5, 0.7$ and (a,b) $\mathrm{Pe}=1$ (c,d) $\mathrm{Pe}=3$ (e,f) $\mathrm{Pe}=5$.}
\label{fig:kn_freq}
\end{figure}
So, we look at statistical properties, averaging over the $M$ independent samples we simulated. In Fig.~\ref{fig:kn_freq}, we report the Knotting frequency (i.e. the fraction of knotted polymers) and the mean gyration radius $\langle R_g \rangle$ as a function of time for $N=300$, $p=0.3, 0.5, 0.7$ and (a,b) $\mathrm{Pe}=1$ (c,d) $\mathrm{Pe}=3$ (e,f) $\mathrm{Pe}=5$. We observe that, after a relatively short amount of time, that we discard (as mentioned in Sec.~\ref{sec:sim}), the average gyration radius fluctuates around a constant value. This indeed ensures us that the ``patological'' conformations, showcased in \cite{vatin2024conformation}, do not appear. Furthermore, the fraction of knotted chains is, after an equilibration time, fluctuating around a constant value. This ensures us that knots are forming and untying at a steady rate.

\subsection{Measurement of knot formation rate and lifetime}
\label{sec:dyn_quant}
\begin{figure}[h!]
\includegraphics[width=0.95\textwidth]{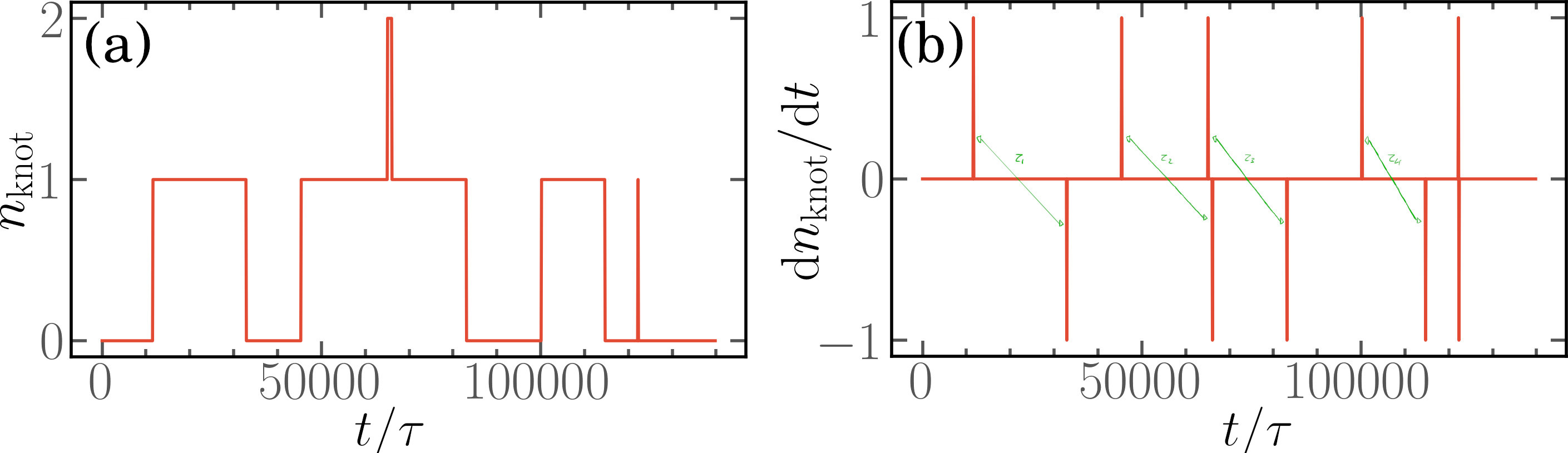}
\caption{a): Number of knots in a particular chain as a function of time. 
b): Time derivative of the number of knots as a function of time. Green arrows indicate the intervals between a positive and a corresponding negative peak of the derivative. For both panels $p = 0.6$, $\mathrm{Pe} = 5$.}
\label{fig:supp1}
\end{figure}
Dynamical information on knots, such as the lifetime, are measured by analyzing the time derivative of the number of knots $\mathrm{d}\,n_\mathrm{knot}/\mathrm{d}t$ along the trajectory; statistics is collected from the $M=$100 independent runs.\\
The rate of formation is measured as the number of positive peaks of $\mathrm{d}\,n_\mathrm{knot}/\mathrm{d}t$ divided by the total simulation time of the trajectory.\\
The knot lifetime is measured as the average time between a positive peak and the corresponding negative peak of $\mathrm{d}\,n_\mathrm{kn}/\mathrm{d}t$. Here, identifying the ``corresponding'' negative peak refers to the identification of the correct tying-untying pair; it is of course relevant only when multiple knots per chain are present. To simplify the analysis, leveraging on the broken symmetry of the system, that causes directional motion of the knots, and on the driven nature of the knot migration, we assume that new knots can not overcome old ones along the chain. We have also never observed such type of event, within the window of parameters considered. As such, when observing two formation events and one untying event (as in Fig.~\ref{fig:supp1}B), we associate the event to the first knot and we measure the lifetime accordingly.\\  
Finally, Kymoknot may wrongly identify a knot in some conditions. Such knots usually last for only one frame of the trajectory (see the last peak in Fig.~\ref{fig:supp1}B). This type of event is clearly not considered as a knot in our analysis, neither in the statistics reported below and in the main text, since it corresponds to a transient configuration of the chain that does not correspond to a physical knot.

\section{Supplementary results}

\subsection{Dynamic of a knot on a partially active ring polymer}
	
\begin{figure}[h!]
\includegraphics[width=0.8\textwidth]{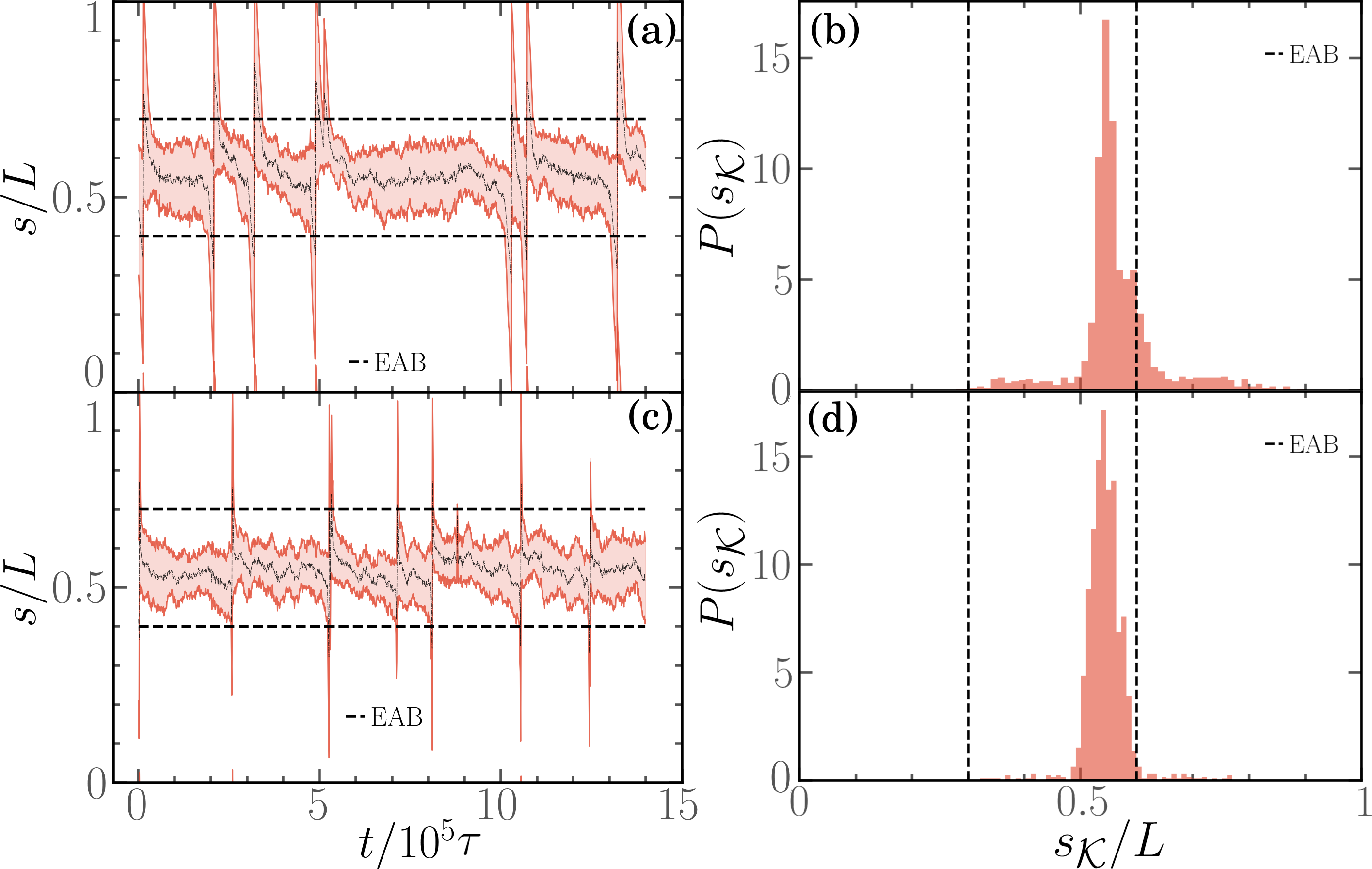}
\caption{Kymograph of the position of the knot (left) and distribution of the knot's center (right) for an active diblock ring polymer with $p = 0.7$ active monomers for $\mathrm{Pe}=1$ (top) and $\mathrm{Pe}=5$ (bottom).}
\label{fig:ring}
\end{figure}

We briefly consider, in this section, the dynamical property of a trefoil ($3_1$) knot in an active/passive diblock copolymer ring. We simulate the ring  polymer employing the model described in Section~\ref{sec:model}, i.e. the same used for the linear chains. Naturally, in this setting the knot can't untie; as such we follow its dynamics in a steady state over a relatively short time span of $\sim 10^3\tau$. We report a few highlights in Fig.~\ref{fig:ring} for a ring of $N=300$, $p=0.7$ and $\mathrm{Pe}=1$ (top), $\mathrm{Pe}=5$ (bottom). Due to the tangential nature of the active force, the knot also travels along the contour and periodically crosses a point, arbitrarily labeled as the beginning/end of the ring.  Interestingly, in both cases the knot spends most of the time in the passive section, highlighted by the two dashed lines, as happens in linear chains. The knot also moves through the active block very fast, due to the high mobility of the active monomers.  

\subsection{Distribution of knot types}
\label{sec:knot_type}
\begin{figure}[h!]
	\includegraphics[width=0.32\textwidth]{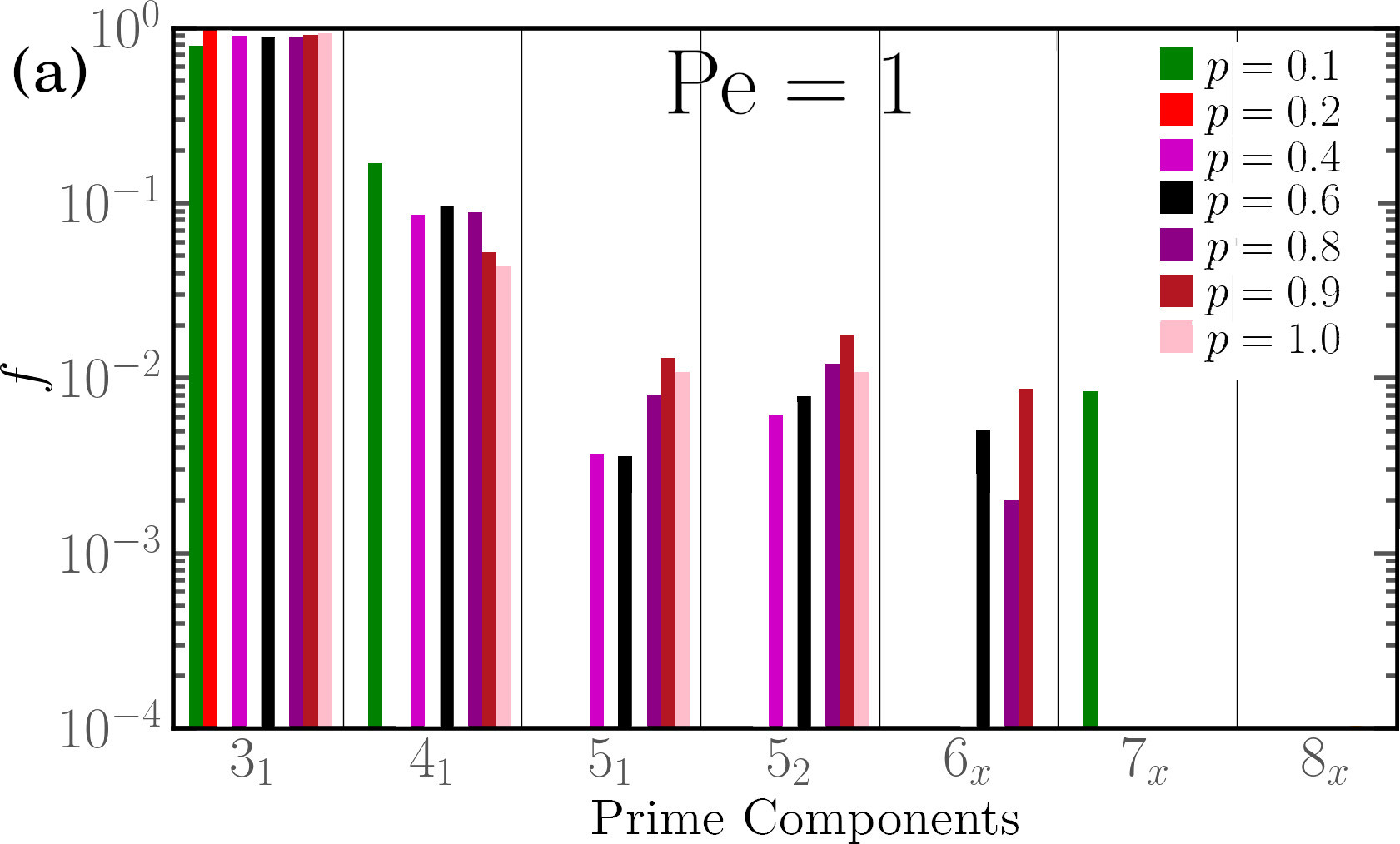}
	\includegraphics[width=0.32\textwidth]{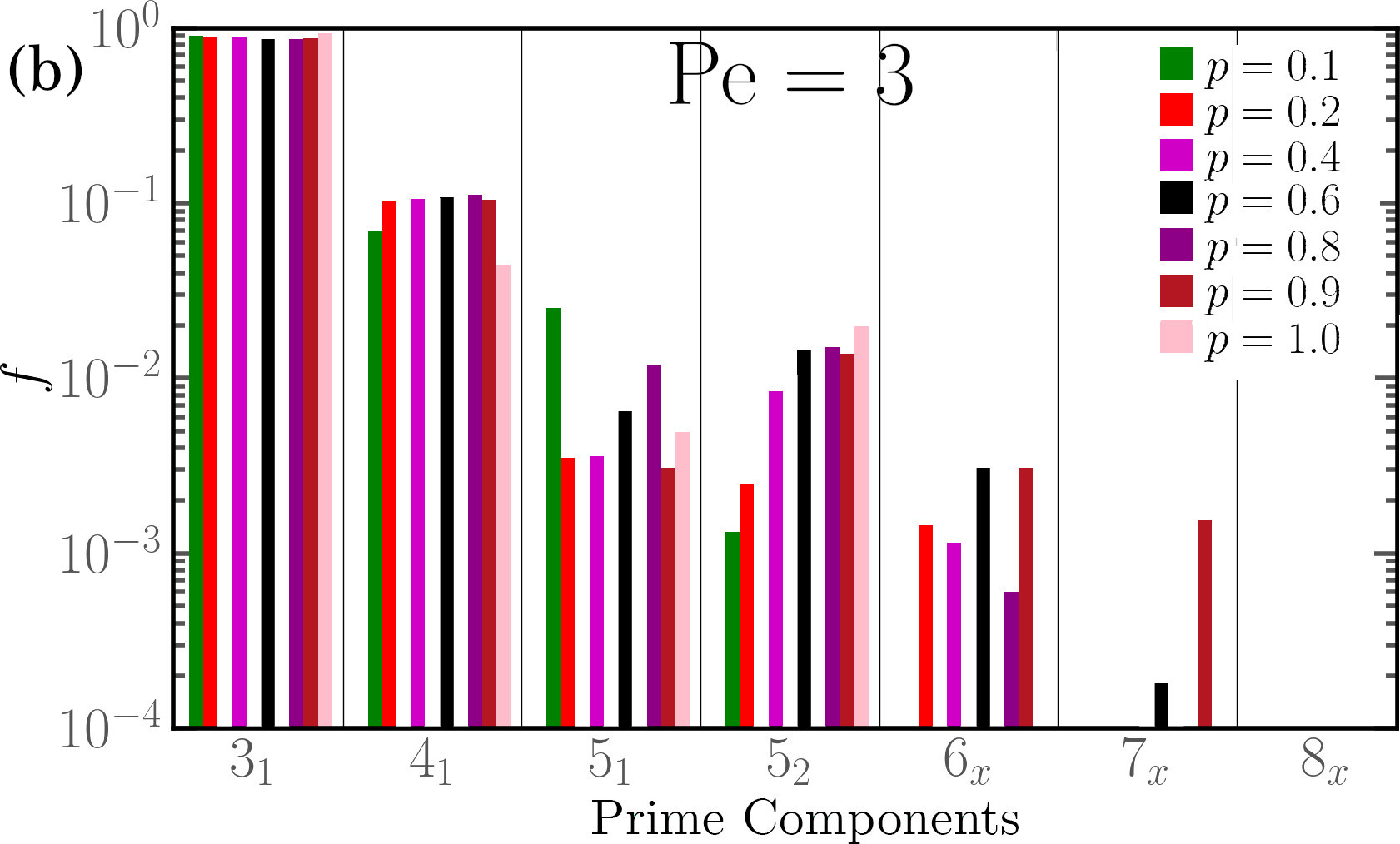}
	\includegraphics[width=0.32\textwidth]{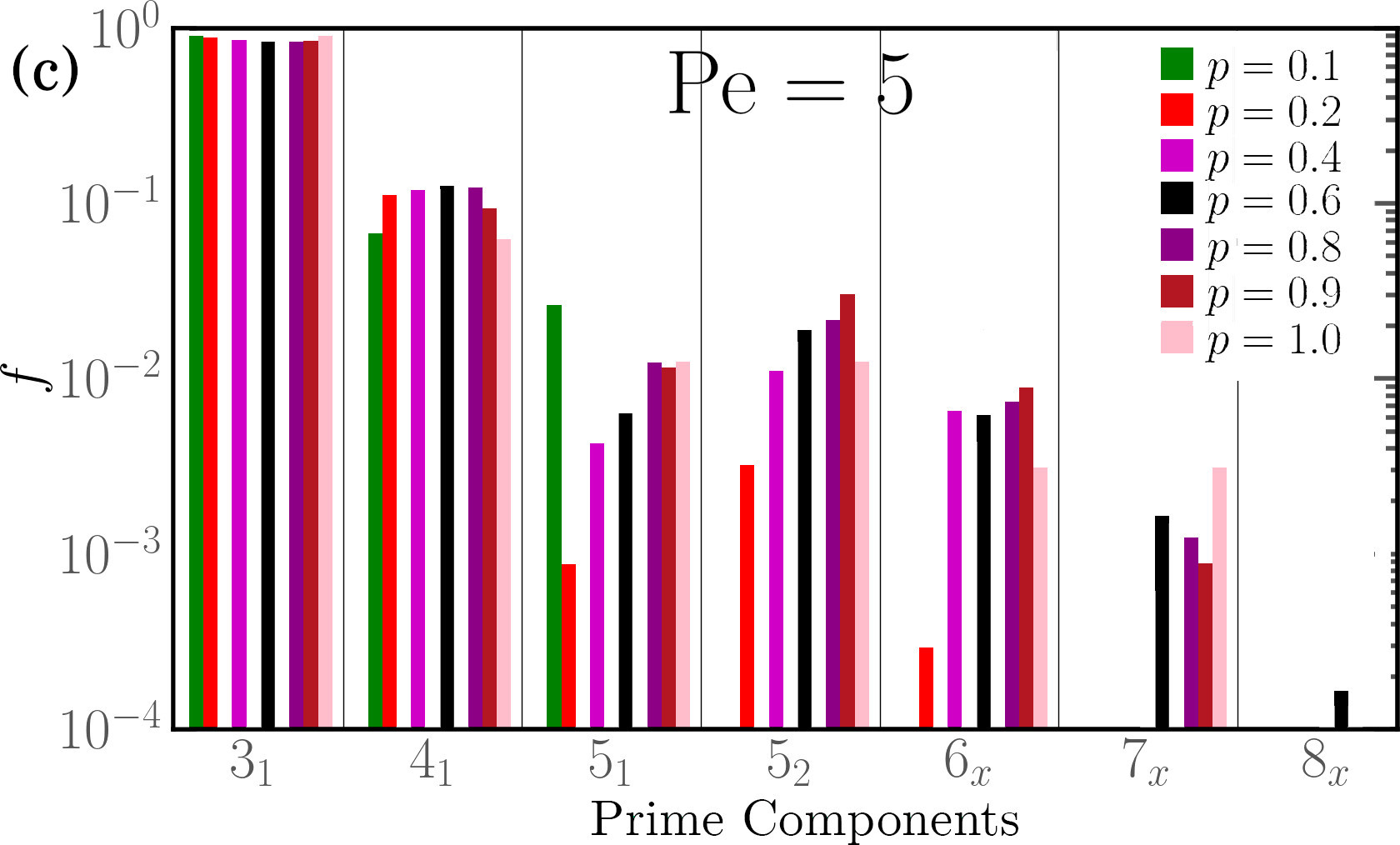}
\caption{Distribution of prime knot types for $N=300$ and $\mathrm{Pe}=1$ (a), $\mathrm{Pe}=3$ (b), $\mathrm{Pe}=5$ (c) for different percentages of active sites $p$.}
\label{fig:kn_dist_1}
\end{figure}

In this section, we present complementary data concerning the distribution of prime knot types for our different sets of parameters. Figs.~\ref{fig:kn_dist_1} show the distribution of prime knot types for $N=300$ and $\mathrm{Pe}=$1,3,5 for different values of $p$. The distribution of knot types is computed by counting the number of occurrences of each prime knot type along the 100 independent trajectories. The results are presented as bar plots, where the height of each bar represents the percentage of occurrence of the corresponding knot type. 
We use the Alexander-Briggs notation to label the knots, where the first number indicates the number of crossings, and the second number is a label indicating the type of knot. From knots presenting 6 to 8 crossings, we use $x$ as subscript label instead of a number in order to include all the knots having a given number of crossings. As mentioned in the main text, the complexity of the knots slighly increases upon increasing $\mathrm{Pe}$; however, the overwhelming majority of the knots are $3_1$ and $4_1$.

\subsection{Distribution of the positions of the center of the knot}
\label{sec:knot_pos}
\begin{figure}[h!]
\includegraphics[width=0.32\textwidth]{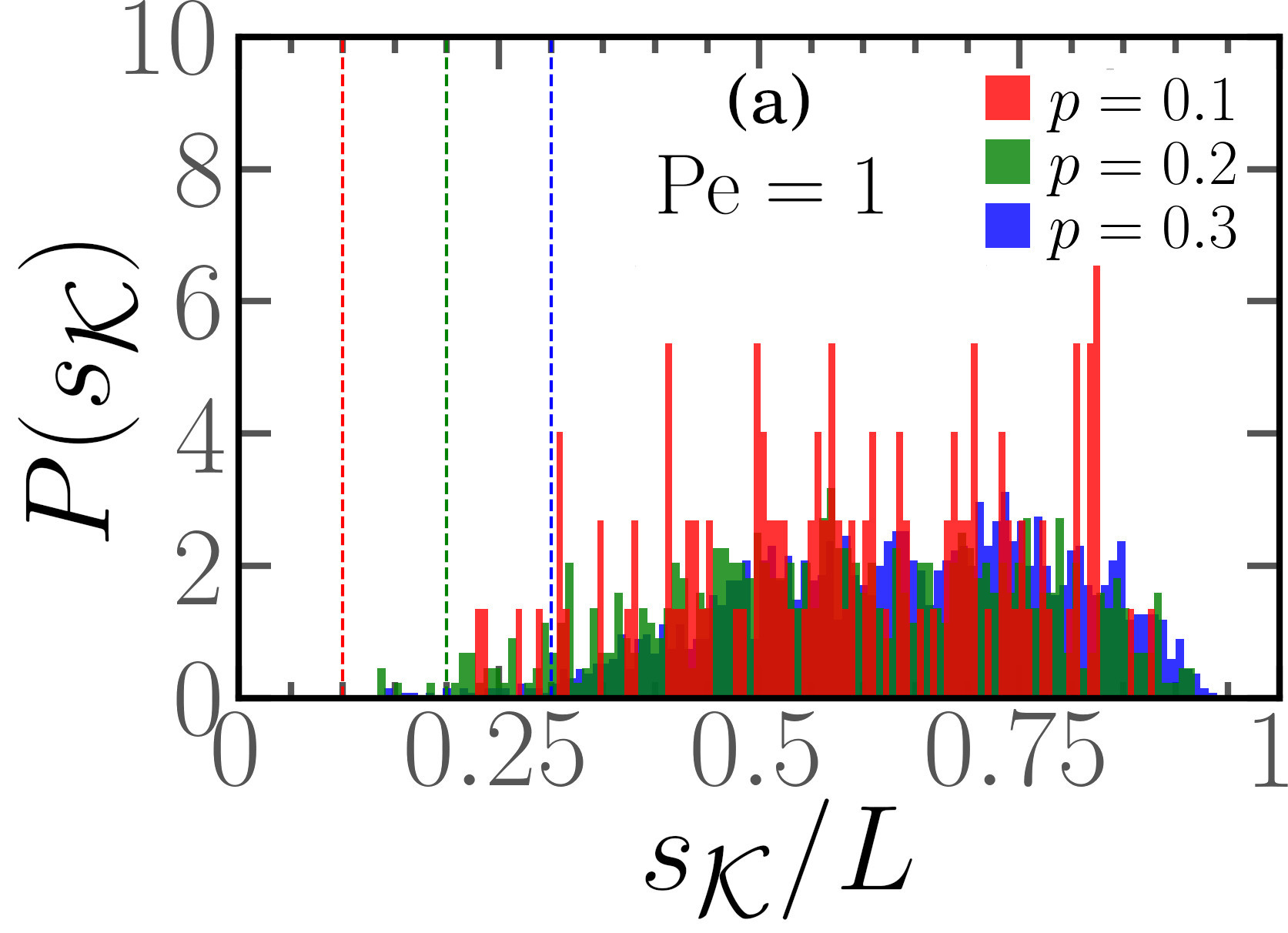}
\includegraphics[width=0.32\textwidth]{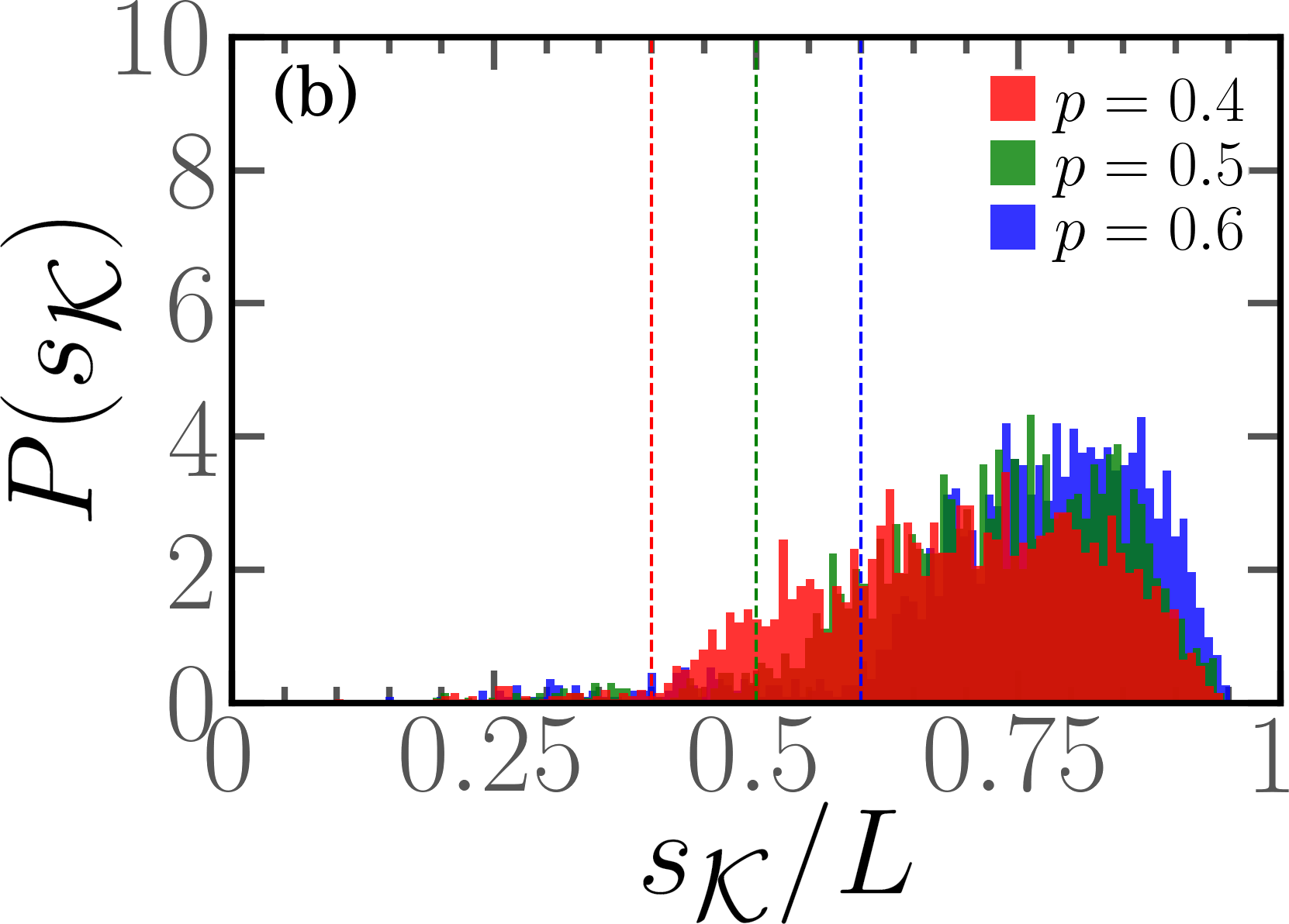}
\includegraphics[width=0.32\textwidth]{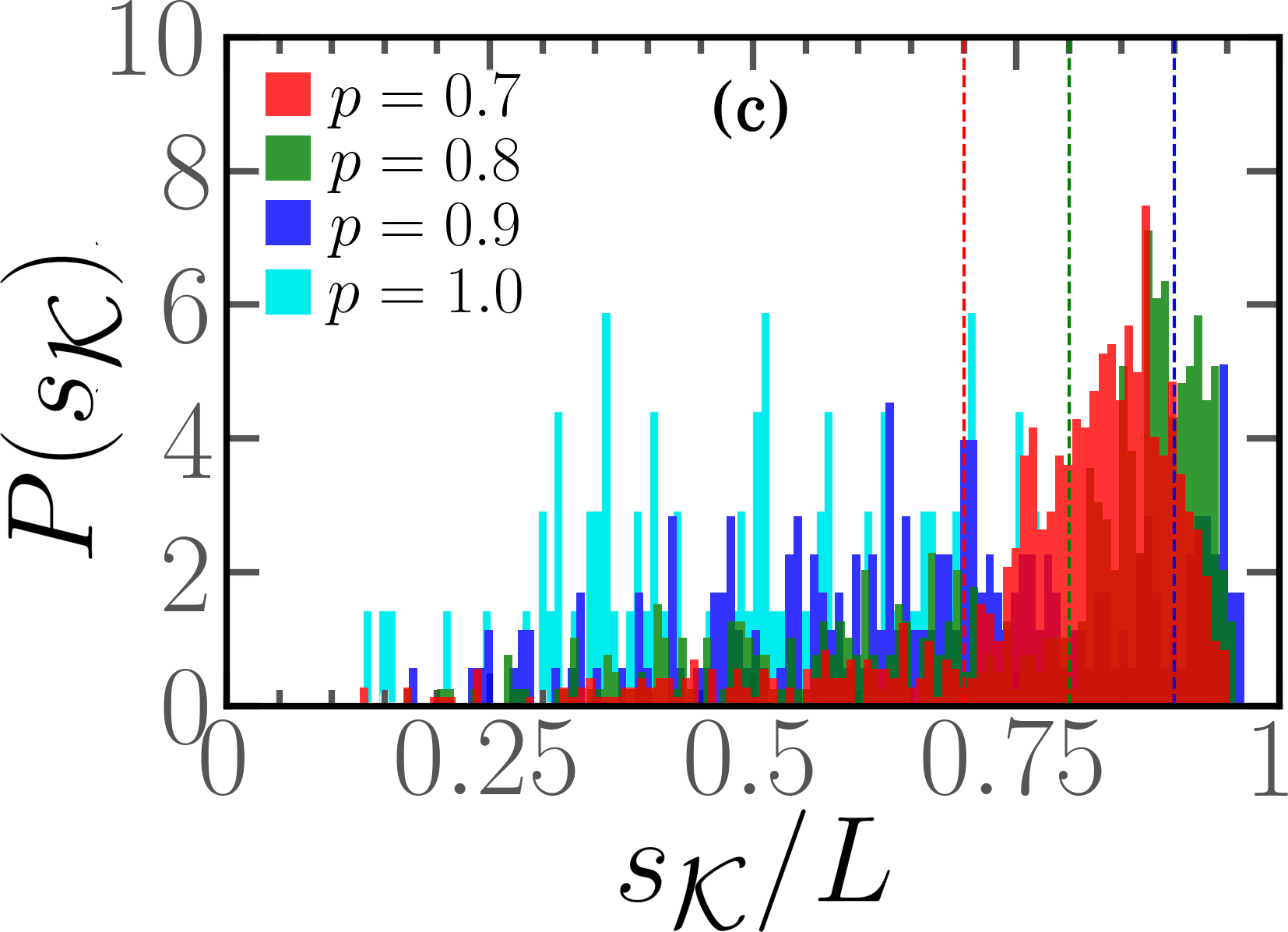}\\
\includegraphics[width=0.32\textwidth]{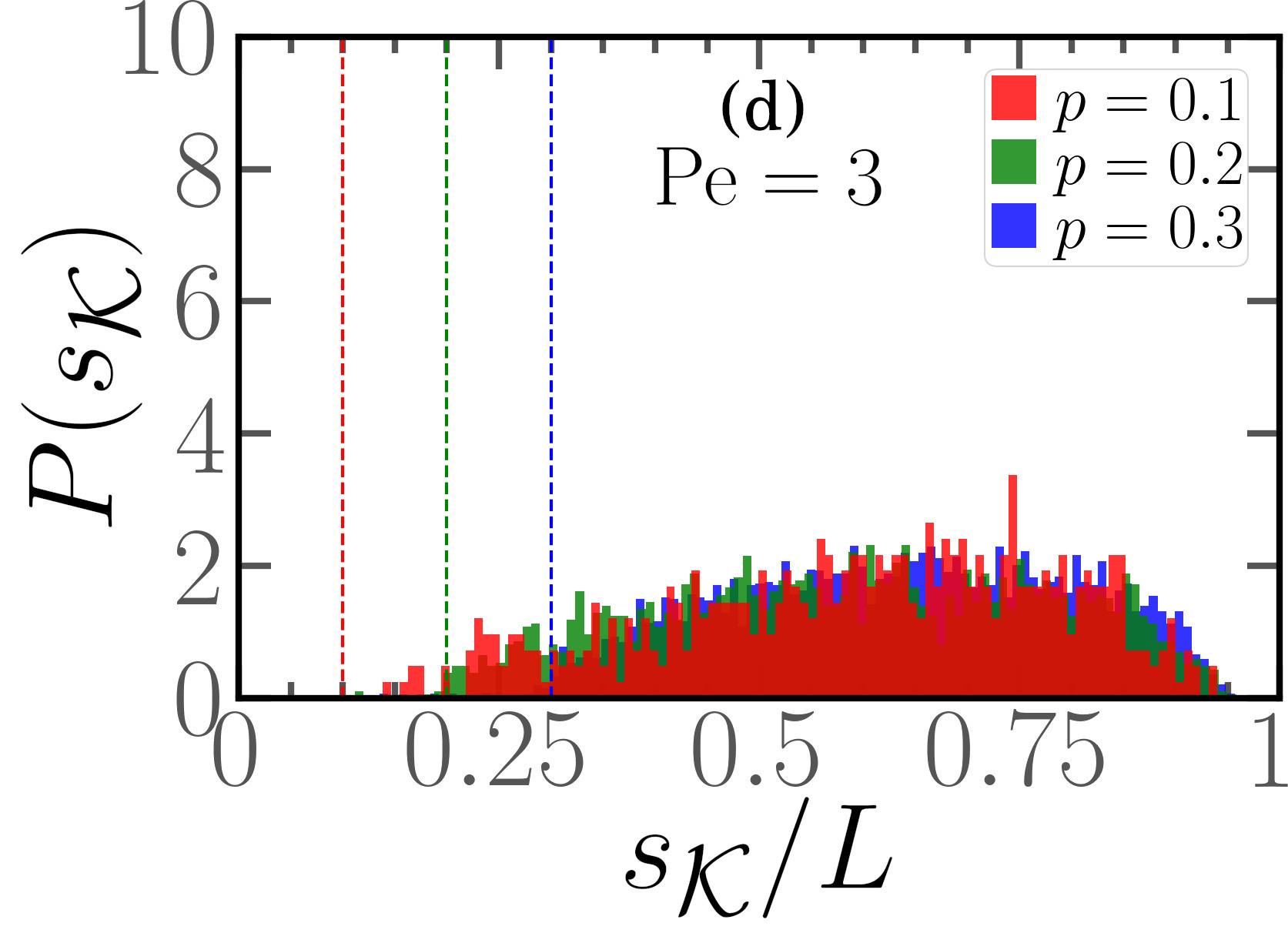}
\includegraphics[width=0.32\textwidth]{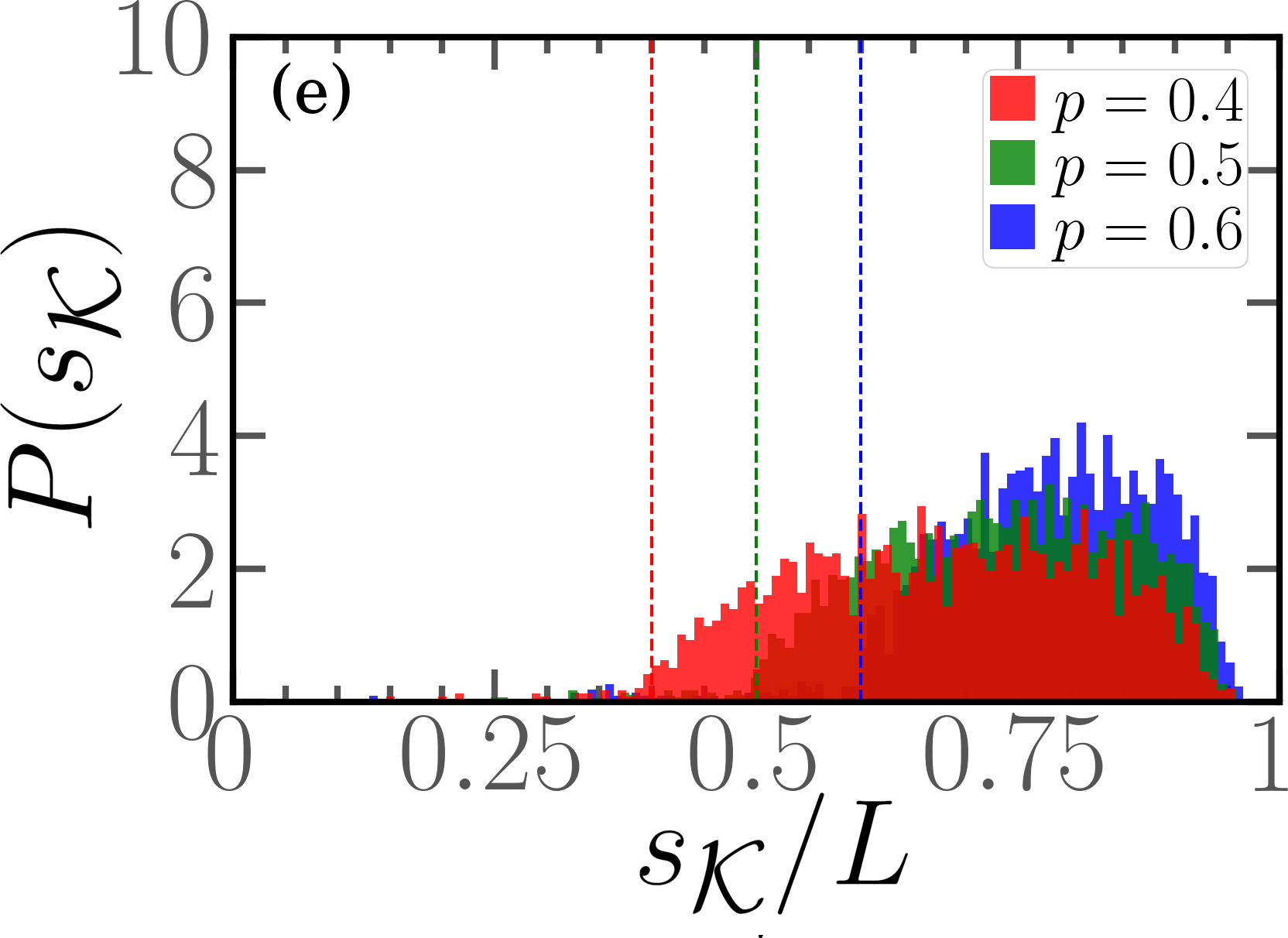}
\includegraphics[width=0.32\textwidth]{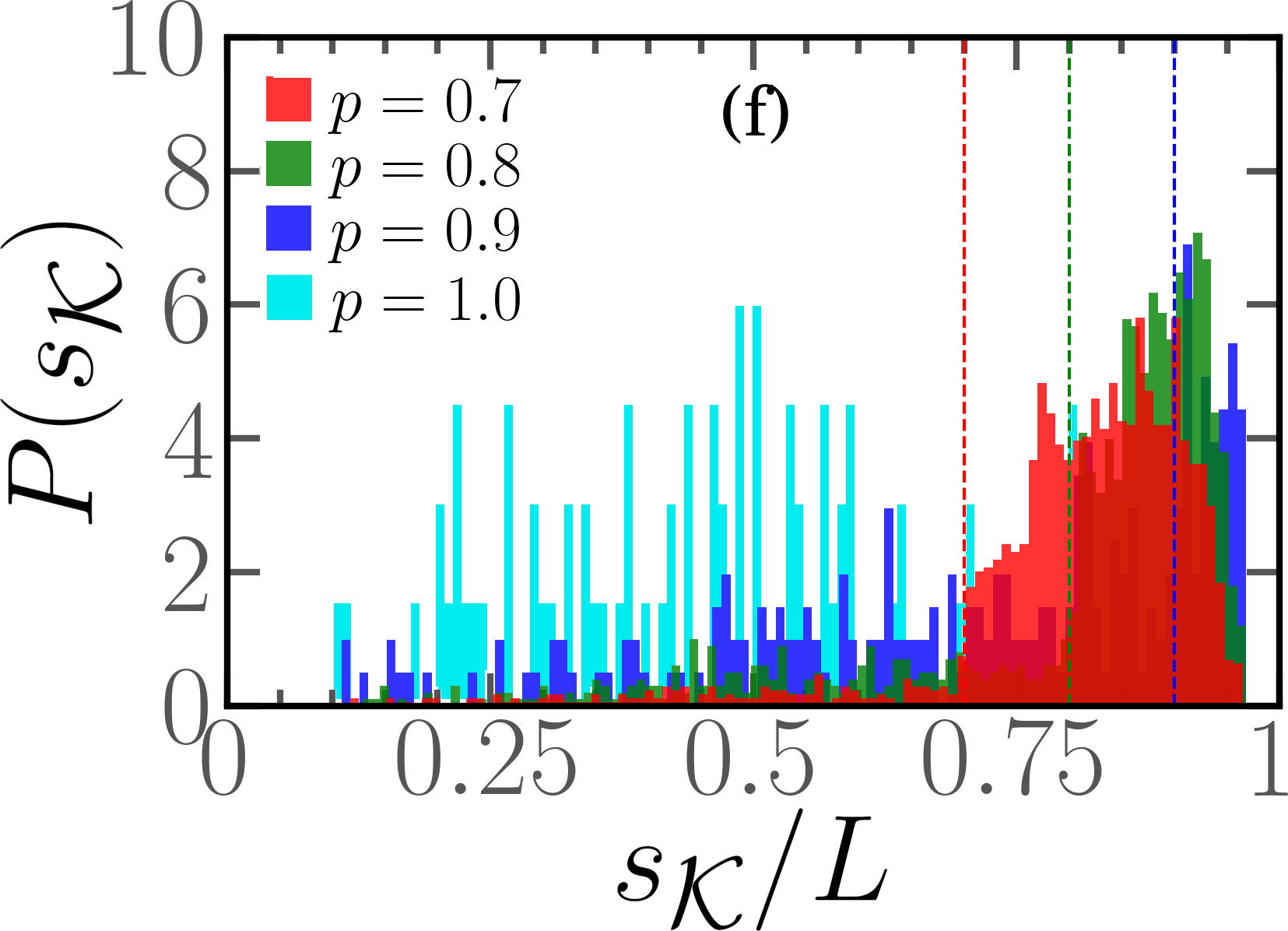}\\
\includegraphics[width=0.32\textwidth]{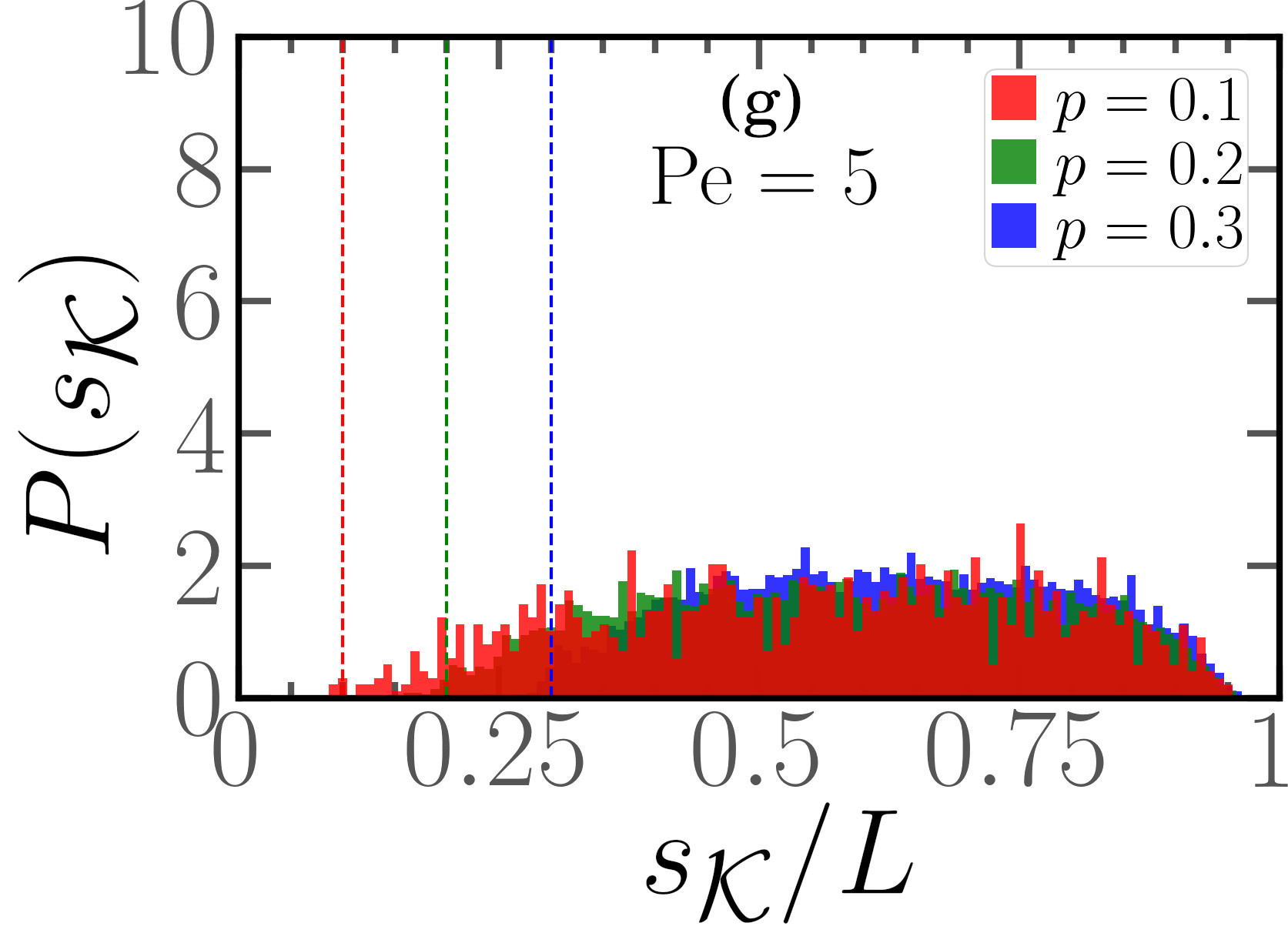}
\includegraphics[width=0.32\textwidth]{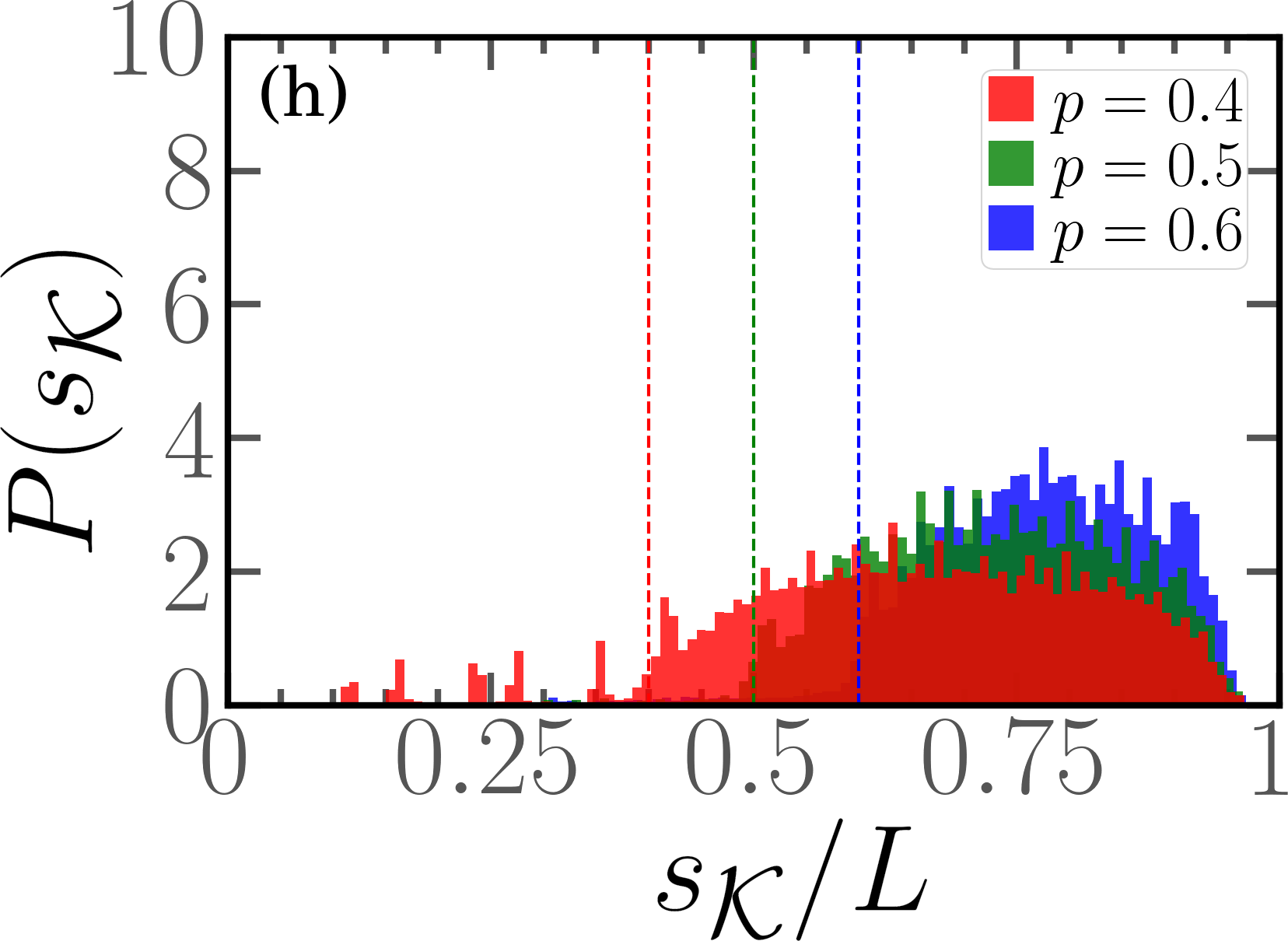}
\includegraphics[width=0.32\textwidth]{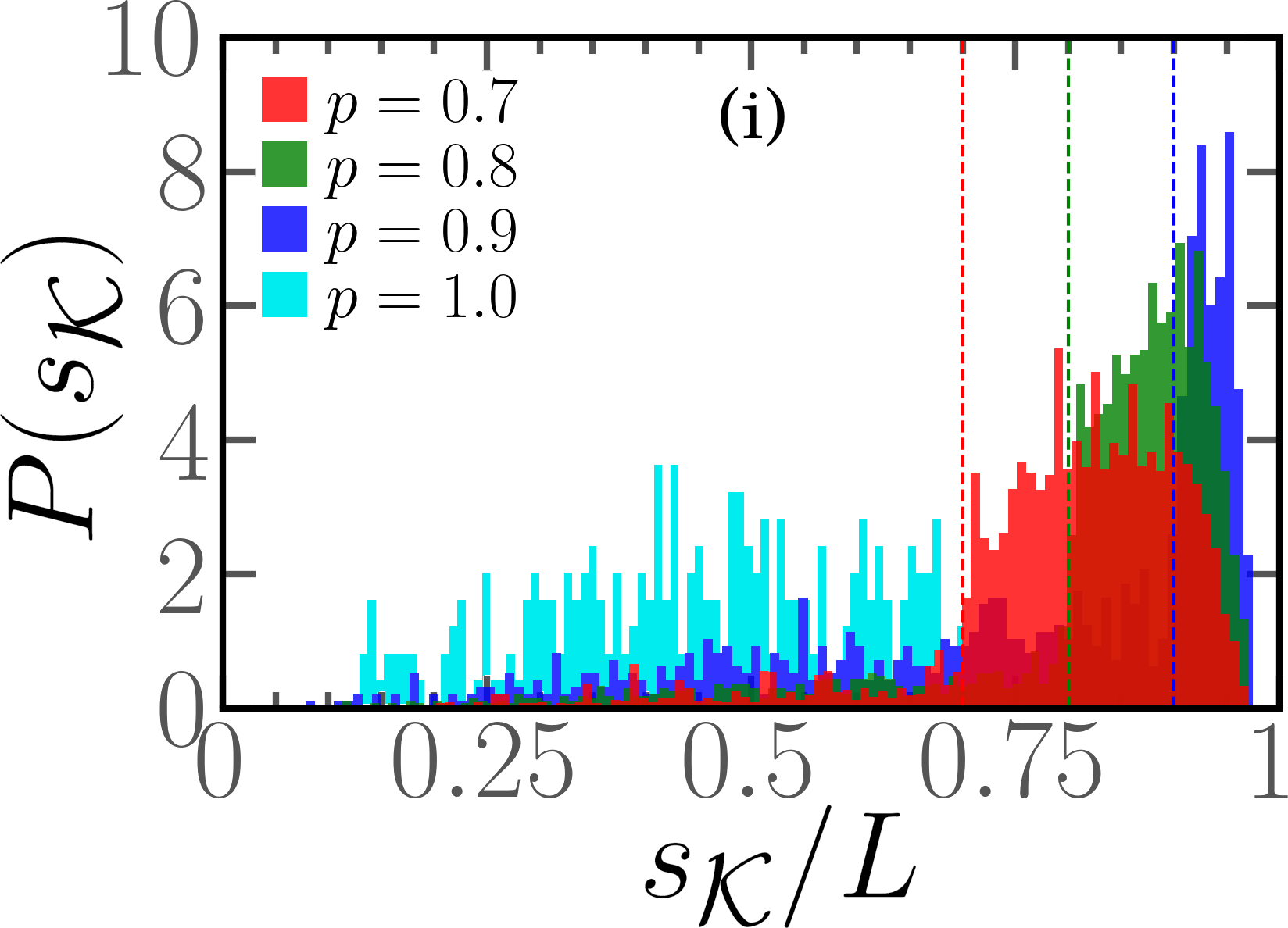}
\caption{Distribution of $s_{\mathcal{K}}$ (prime components) for $N=300$ and $\mathrm{Pe}=1$ (a-c), $\mathrm{Pe}=3$ (d-f), $\mathrm{Pe}=5$ (g-i) at different values of $p$.}
\label{fig:kn_pos}
\end{figure}

In this section, we present complementary data concerning the distribution of the positions of the center of individual prime components along the polymer chain for different values of $\mathrm{Pe}$ and $p$. The distributions are reported in Fig.~\ref{fig:kn_pos}
A vertical dashed line (with the same color as the corresponding data) is drawn in order to highlight the position of the end of the active block (EAB); one can appreciate that knots are mainly located in the passive block up to $p=0.7-0.8$.  

\subsection{Distribution of the lengths of the knots}
\label{sec:knot_length}
\begin{figure}[h!]
\includegraphics[width=0.32\textwidth]{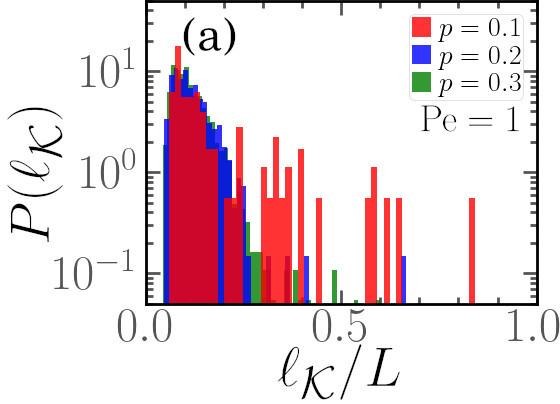}
\includegraphics[width=0.32\textwidth]{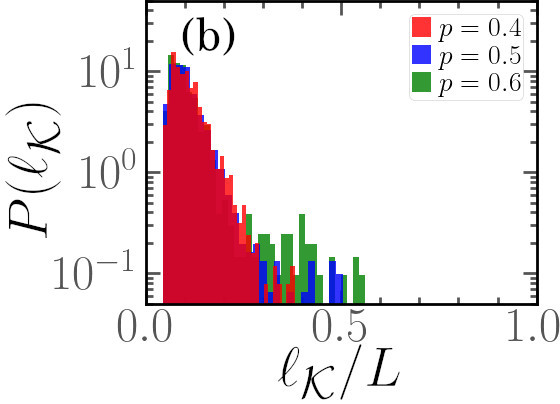}
\includegraphics[width=0.32\textwidth]{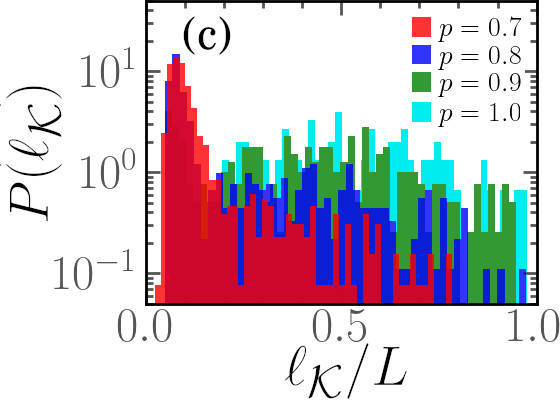}\\
\includegraphics[width=0.32\textwidth]{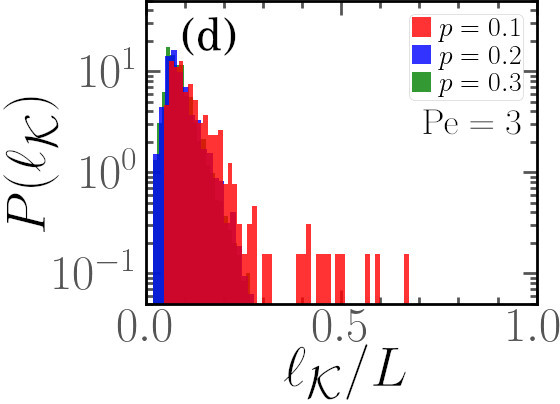}
\includegraphics[width=0.32\textwidth]{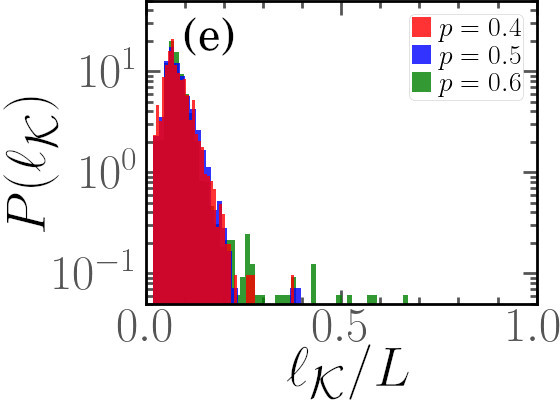}
\includegraphics[width=0.32\textwidth]{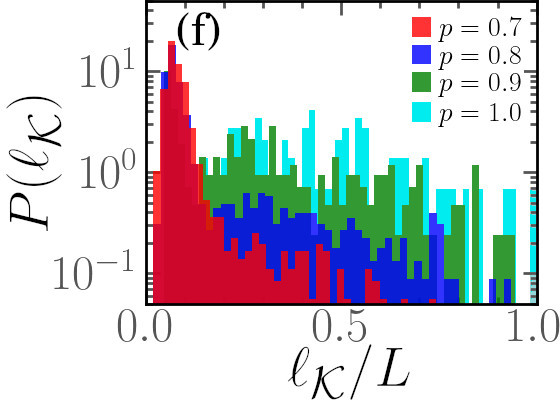}\\
\includegraphics[width=0.32\textwidth]{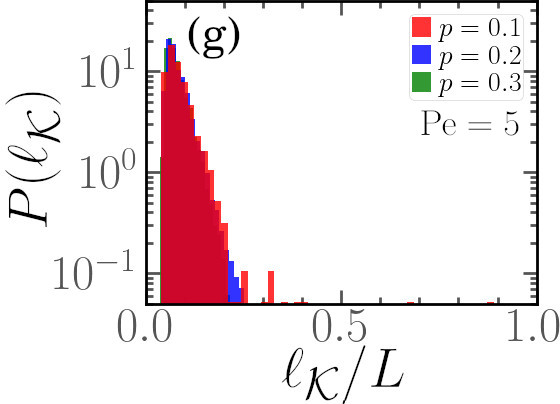}
\includegraphics[width=0.32\textwidth]{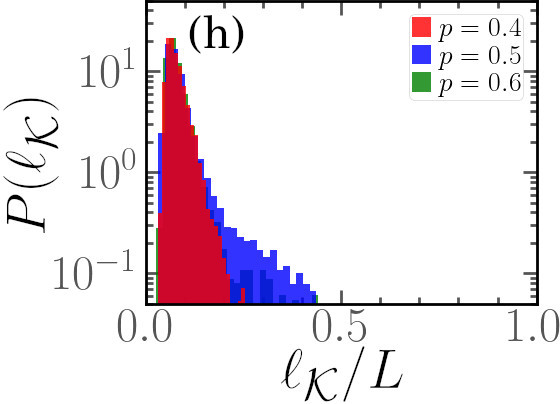}
\includegraphics[width=0.32\textwidth]{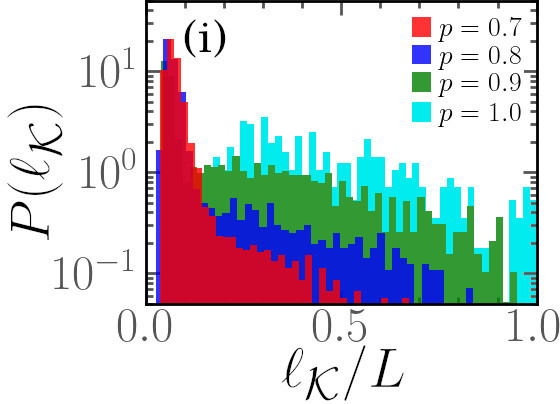}
\caption{Distribution of $\ell_{\mathcal{K}}$ (prime components) for $N=300$ and $\mathrm{Pe}=1$ (a-c), $\mathrm{Pe}=3$ (d-f), $\mathrm{Pe}=5$ (g-i) at different values of $p$.}
\label{fig:kn_length}
\end{figure}

This section presents complementary data concerning the distribution of the lengths of individual prime knots for different values of $\mathrm{Pe}$ and $p$. The distributions of knot lengths are presented as histograms in Figs.~\ref{fig:kn_length}. The length of a knot is defined as the contour length of the knotted region, as identified by Kymoknot.  It can bee seen that the majority of the knots are made by approximately $10\%$ of the monomers of the chain, i.e. about 30 monomers. As reported in the main text, upon increasing $p$ the distributions develop a second tail at large values of $\ell_\mathrm{kn}$, corresponding to the cases where the knots, still in the active section, are delocalized and spread over the entire active section of the chain.

\subsection{Correlation the knots' length and position}
\label{sec:correlation}
\begin{figure}[h!]
\includegraphics[width=0.9\textwidth]{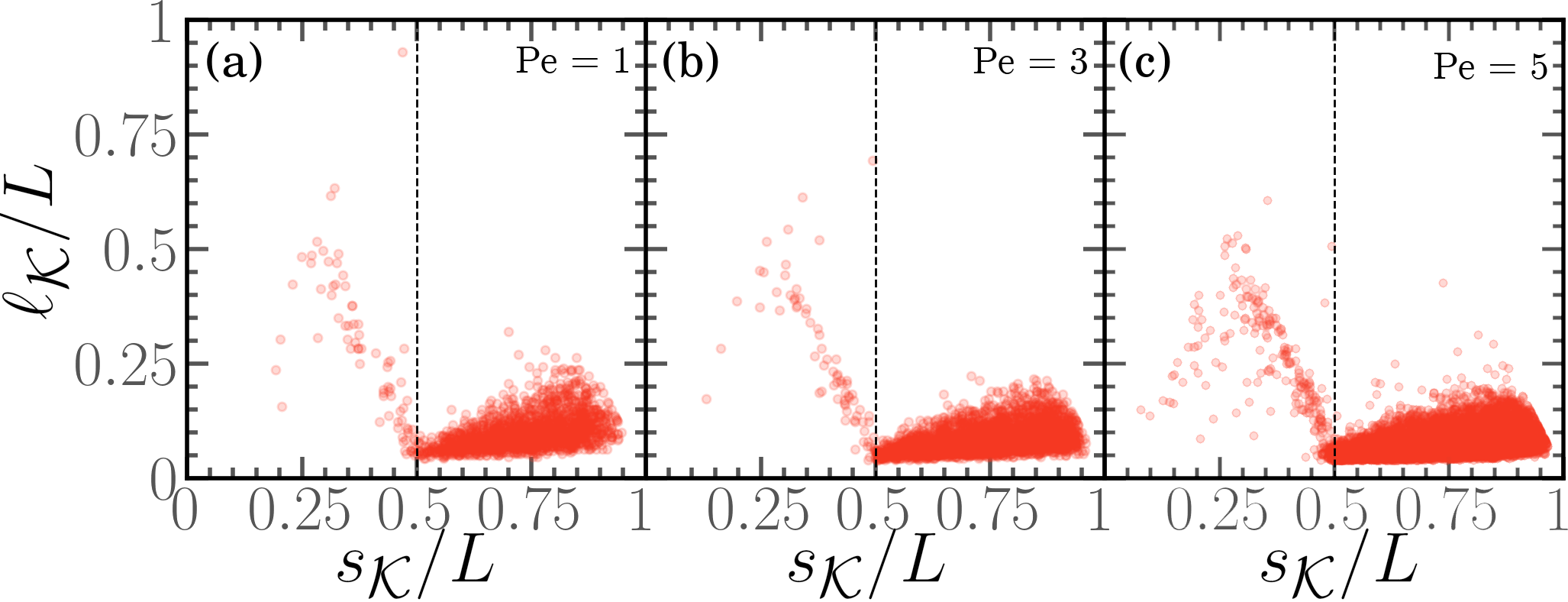}
\caption{Correlation between the reduced knot length $\ell_{\mathcal{K}}/L$ and the reduced position of their center $s_{\mathcal{K}}/L$, for $N=300$ and $\mathrm{Pe}=1$, $\mathrm{Pe}=3$, $\mathrm{Pe}=5$ for $p=0.5$.}
\label{fig:kn_corr_1}
\end{figure}

\begin{figure}[h!]
\includegraphics[width=0.95\textwidth]{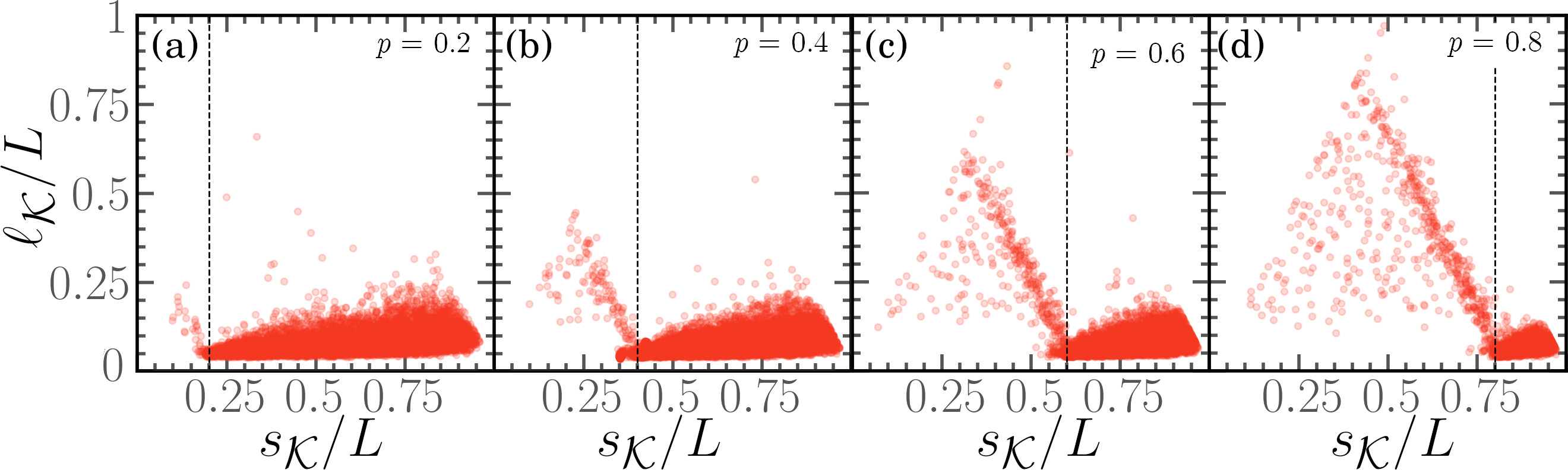}
\caption{Correlation between the reduced knot length $\ell_{\mathcal{K}}/L$ and the reduced position of their center $s_{\mathcal{K}}/L$, for $N=300$ and $\mathrm{Pe}=5$ for $p=0.2$, $p=0.4$, $p=0.6$, $p=0.8$.}
\label{fig:kn_corr_2}
\end{figure}

In this section, we present complementary data concerning the correlation between the length of the knots and the position of their center along the polymer chain for our different sets of parameters. The results are presented as scatter plots in Figs.~\ref{fig:kn_corr_1} to \ref{fig:kn_corr_2}. 
In Fig.~\ref{fig:kn_corr_1}, we report scatter plots of $s_{\mathcal{K}}/L$ versus $\ell_{\mathcal{K}}/L$ at  fixed $p=$0.5 and different values of $\mathrm{Pe}$. We observe that, in all three cases, the knot in the active region is de-localized; it can be even fully spread over the region ($\ell_{\mathcal{K}}/L \simeq p$). The scattered points have also a marked triangular pattern in the active region, that indicates how the knot shrinks, while approaching the active passive boundary. On the other hand,  triangular, asymmetric, shape of the data in the passive region quantifies the localisation process and suggests, as also happens in passive polymers, that the knot swells before untying.
Finally, in Fig.~\ref{fig:kn_corr_2}, we report scatter plots of $s_{\mathcal{K}}/L$ versus $\ell_{\mathcal{K}}/L$ at  fixed $\mathrm{Pe}=$5 and different values of $p$. Also these plots confirm the picture given so far; interestingly, for large values of $p$ one can find more frequently relatively small knots in the active section, occurrence that is rarer for small values of $p$.

\end{document}